\documentclass[twocolumn,10pt]{IEEEtran}
\IEEEoverridecommandlockouts
\UseRawInputEncoding
\usepackage{color,graphicx,epsfig,bbm,cite,array,amsfonts,graphics,amsmath,psfrag,subfigure,wrapfig,amssymb,dsfont,floatflt}
\usepackage[justification=centering]{caption}

\begin{document}
\title{High Fidelity RF Clutter Modeling and Simulation}
\author{\IEEEauthorblockN{Sandeep Gogineni$^{1*}$, Joseph R. Guerci$^{1}$, Hoan K. Nguyen$^{1}$, Jameson S. Bergin$^{1}$,\\ David R. Kirk$^{1}$, Brian C. Watson$^{1}$, Muralidhar Rangaswamy$^{2}$\\}\thanks{$*$Corresponding author email: sgogineni@islinc.com. The opinions and statements in this work are the authors' own and does not constitute any explicit or implicit endorsement by the U.S. Department of Defense. Drs. Rangaswamy and Gogineni were supported by the Air Force Office of Scientific Research under projects 20RYCOR051 and 20RYCOR052, respectively.}
\IEEEauthorblockA{$^{1}$ Information Systems Laboratories, Inc., USA\\}
\IEEEauthorblockA{$^{2}$ Air Force Research Laboratory, Wright Patterson Air Force Base, OH, USA\\}}
\maketitle

\begin{abstract}
In this paper, we present a tutorial overview of state-of-the-art radio frequency (RF) clutter modeling and simulation (M\&S) techniques. Traditional statistical approximation based methods will be reviewed followed by more accurate physics-based stochastic transfer function clutter models that facilitate site-specific simulations anywhere on earth. The various factors that go into the computation of these transfer functions will be presented, followed by several examples across multiple RF applications. Finally, we introduce a radar challenge dataset generated using these tools that can enable testing and benchmarking of all cognitive radar algorithms and techniques.\\

{\IEEEkeywordsname : High-Fidelity, Physics-Based, Modeling and Simulation, Radio Frequency, Cognitive Radar, Stochastic Transfer Function, Green's Function Impulse Response}
\end{abstract}
\section{Introduction}
Radio frequency (RF) signals are used in a multitude of defense, commercial, and civilian applications that are critical to the safety and security of mankind. Most of the RF applications like Radar include target detection, localization, and tracking in the presence of intentional and unintentional interference. In this paper, although we focus on radar applications, the techniques presented herein are relevant to all RF applications. In a radar system, an RF transmitter sends out signals to illuminate a scene of surveillance to infer about the scene and the targets present based on the echo signal measured at the receiver. In an ideal world without any interfering signals, accomplishing these tasks is fairly trivial. However, in a practical setting, the RF signals at the receiver are almost always corrupted by interfering signals. A major source of interference is reflections off ground clutter which are highly dependent on the terrain present in the illuminated scene.

Targets of interest can be obscured by these ground clutter reflections and this interference is even more prominent when the radar system is flying in the air, looking at ground targets. Therefore, the development of any new radar technique is heavily dependent on accurately modeling these ground clutter reflections. The models are critical not only in the development stage but also in the testing and evaluation phase. There is a scarcity of publicly available measured data for RF applications. The measured data is expensive to collect and limited to very specific scenarios. Even when collected, the data is sensitive in nature and not readily available to test new algorithms and techniques. Therefore, most of the radar research, development, and testing relies upon accurately modeling and simulating the data.

Traditional approach to clutter modeling treats the clutter returns to be random vectors with unknown covariance matrices \cite{GuerciBook}--\nocite{Guerci2}\nocite{Guerci3}\nocite{KaySTSP}\cite{PillaiGuerci}. Initially, the covariance matrices were assumed be constant for any given scenario since traditional radar systems always transmitted fixed waveforms. With the advent of cognitive radar systems that are capable of adapting transmit waveforms, these models have been modified to treat the covariance matrices as a function of the transmit waveform. However, even with this change, these traditional models that have been used for several decades are still a statistical approximation as they essentially treat the clutter signals to be fully random in nature.

In reality, the clutter signals measured at any scene always include a deterministic component that is dependent on the physical features of the scene that has been illuminated. For example, the mountains, rivers, lakes, etc within a scene do not move and hence, if we collect radar data over the same scene on multiple days, we will have a common deterministic component to these measurements. There will be a random component as well, owing to other variations such as swaying of trees, waves on the water surface, etc. In the absence of any information about the fixed features present in the scene, the random models described in the previous paragraph can be used. However, when we have access to real-world environmental databases, an M\&S tool must be able to faithfully replicate these site-specific features. Inspired by this, in \cite{Guerci}, an alternate approach to clutter modeling using a "stochastic transfer function" (Green's function impulse response in the time domain) approach has been presented for this problem. This results in a fundamental physics based scattering model that can be used to accurately simulate RF data.

In Section II, the traditional covariance-based model will be described in more detail, followed by the stochastic transfer function model in Section III. In Section IV, we present several RF examples using this new approach to M\&S. Finally, in Section V, we introduce a radar challenge dataset generated using the site-specific stochastic transfer function model that can be used by the readers to test and benchmark all cognitive radar algorithms and techniques. Some examples of the algorithms that can be bench-marked using this dataset include adaptive waveform design and channel estimation. Conclusions are presented in Section VI.

\section{Traditional Covariance-based Statistical Model}
Traditional space-time adaptive processing literature treats radar returns from ground clutter as completely random with a pre-described probability distribution \cite{GuerciBook2}, \cite{Ward}. Considerable history underlies the exposition of \cite{Ward}. Early work in this direction involved the collection and analysis of experimental data \cite{georgenrl}, \cite{trunk}, which attempted to fit two parameter families of distributions to describe the heavy tailed behavior of clutter returns corresponding to high resolution radar for false alarm regulation. In an attempt to account for the pulse-to-pulse correlation as well as the first order probability density function, endogenous and exogenous clutter models were developed \cite{FarinaWeibull}--\nocite{GLi89}\nocite{Conte}\nocite{Rangaswamy93}\nocite{Rangaswamy95}\cite{Sangston94}. The corresponding problem for coherent processing in Gaussian clutter received much attention from the 1950s \cite{ReedIRE}--\nocite{Howell}\nocite{Applebaum}\nocite{Widrow67}\cite{RMB74}. Extensions of these treatises to account for CFAR behavior of the underlying adaptive processor were undertaken in \cite{Kelly86}--\nocite{Kelly89}\nocite{RFKN}\cite{ScharfCFAR}. All of these treatises use statistical approximations for clutter as described herein.

\begin{figure}[htbp!]
    \centering
    \includegraphics[scale=0.7]{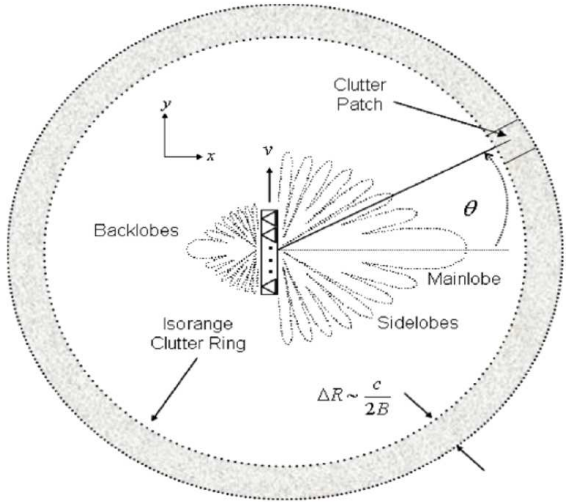}
    \caption{Traditional covariance-based clutter model: Illustration of the monostatic iso-clutter range cell observed from a stand-off airborne radar}
    \label{trad}
    \end{figure}

In this section, we provide a brief overview of these traditional approaches to clutter modeling. These models essentially treat clutter as an additive “colored noise” process with various approximate probability distribution models \cite{billingsley}. Fig. \ref{trad} depicts the basic clutter physics model under consideration for a generally monostatic airborne moving target indicator radar (for both airborne and ground based targets, AMTI and/or GMTI)—although the approach developed can be easily generalized to other configurations such as bi/multi-statics.

As can be seen from Fig. \ref{trad}, the clutter returns corresponding to a particular range bin of interest can be expressed as a weighted summation of the returns from individual clutter patches present in that ring. Let there be $N_\mathrm{c}$ clutter patches in a given range bin of interest. Then the clutter response corresponding to that range bin can be expressed as
\begin{equation}
\boldsymbol{x_\mathrm{c}}=\sum_{i=1}^{N_\mathrm{c}}{\gamma_i\boldsymbol{v}_i},
\end{equation}
where $\boldsymbol{x_\mathrm{c}}$ is the complex valued $NM$ dimensional space-time total clutter return for a given range bin associated with $N$ spatial and $M$ temporal receive degrees-of-freedom (DoFs). $\boldsymbol{v}_i$ is the space-time steering vector to the $i^{\mathrm{th}}$ clutter patch and is a Kronecker product of the temporal and spatial steering vectors.

While the steering vectors are deterministic, the traditional clutter models treat the complex scalar reflectivity corresponding to each patch to be zero-mean random variables. These variables $\gamma_i$ denote the amplitude corresponding to the $i^{th}$ clutter patch and they are a function of the intrinsic clutter reflectivity and the transmit-receive antenna patterns. Given this model, the associated space-time clutter covariance matrix can be expressed as
\begin{equation}
\mathrm{E}\left\{\boldsymbol{x_c}\boldsymbol{x_c}^H\right\}=\sum_{i=1}^{N_\mathrm{c}}\sum_{j=1}^{N_\mathrm{c}}{\mathrm{E}\left\{\gamma_i \gamma_j^*\right\}\boldsymbol{v}_i\boldsymbol{v}_j^H},
\end{equation}
where $\mathrm{E}\left\{.\right\}$ denotes expectation operation. Under the assumption that these coefficients corresponding to different clutter patches are independent, we can express the clutter covariance matrix as
\begin{equation}
\mathrm{E}\left\{\boldsymbol{x_c}\boldsymbol{x_c}^H\right\}=\sum_{i=1}^{N_\mathrm{c}}{G_i\boldsymbol{v}_i\boldsymbol{v}_j^H}.
\end{equation}

While this traditional approach has been used for the past several decades, it is essentially a statistical approximation and has not been derived from physics like the model described in the next section. Additionally, all the transmit degrees of freedom (DOFs) have been collapsed into a single complex reflectivity random variable and hence appear non-linearly and indirectly in the above equations. Also, under this traditional clutter model, as we can see the clutter returns are independent of the transmitted radar waveform. While this assumption was acceptable for conventional radar systems that transmit a fixed waveform, it is highly unrealistic to assume this model for more modern cognitive radar systems that continuously adapt the transmit waveform to match the operating environment. An important implication of bringing to bear the transmit degrees of freedom is the generation of signal dependent interference. In classical space-time adaptive radar processing, the problem is one of designing a finite impulse response (FIR) filter to adapt to an unknown interference covariance matrix. However, in a given adaptation window, the covariance matrix albeit unknown remains fixed. This fact makes it possible to collect replicas of training data sharing the same covariance structure to form an estimate of the covariance matrix. However, when the transmit degrees of freedom are brought to bear, the observed covariance matrix on receive is a non-linear function of the transmit signal. As a consequence each realization of training data now corresponds to a different covariance matrix. Therefore using such training data for covariance matrix estimation yields an inaccurate estimate of the covariance matrix at best and a singular estimate of the covariance matrix at worst, thereby seriously degrading the performance/implementation of the adaptive processor. Therefore, an advanced clutter modeling approach that can capture the signal dependent nature of ground clutter returns is required. We shall describe one such modeling approach in the next section.

\begin{figure*}[htbp!]
    \centering
    \includegraphics[width=\textwidth,height=10cm]{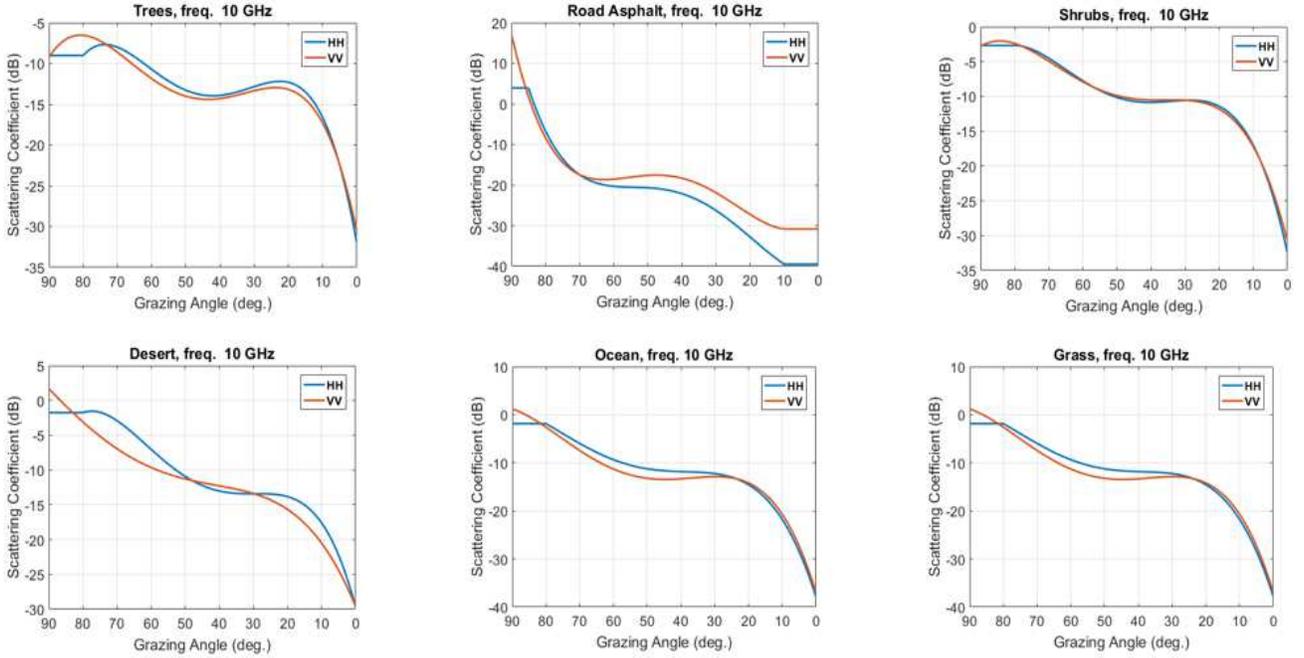}
    \caption{The polarimetric scattering coefficient as a function of grazing angle for different landcover types at X-Band.}
    \label{scatpol}
    \end{figure*}

\section{Stochastic Transfer Function Model}
\begin{figure}[htbp!]
    \centering
    \includegraphics[scale=1.5]{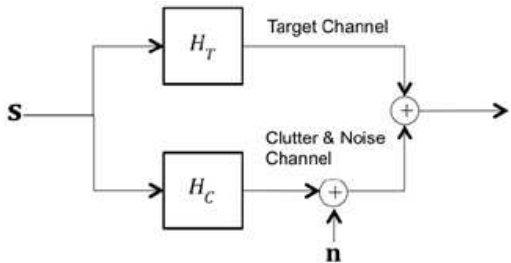}
    \caption{Illustration of the stochastic transfer function model}
    \label{stf_model}
    \end{figure}

Contrary to the covariance-based model, the stochastic transfer function model treats the radar measurements according to the block diagram described in Fig. \ref{stf_model}. This is an accurate representation of the signals since the radar electromagnetic signal travels through the channel interacting with the different components present in the channel in a linear fashion as described by Maxwell's equations. Due to the linear nature of these interactions, the overall channel impact can be represented using an impulse response (Green's Functions impulse response) in the time-domain or the corresponding stochastic transfer function in the frequency domain. The stochastic aspect of the transfer function comes from the random components present in the scene such as intrinsic clutter motion. Note that this new approach to clutter modeling in Fig. \ref{stf_model} separates the radar data into target and clutter channels. The main focus of this paper is the clutter channel.

Let $\boldsymbol{s(n)}$ denote the transmit waveform and $\boldsymbol{h_c(n)}$, $\boldsymbol{h_t(n)}$ denote the Green's function impulse response for the clutter and target channels respectively. Additionally, let $\boldsymbol{n(n)}$ represent the additive thermal noise. Then, the measurements at the radar receiver for time instant $n$ can be represented as
\begin{equation}
\boldsymbol{y(n)}=\boldsymbol{h_c(n)}\circledast\boldsymbol{s(n)}+\boldsymbol{h_t(n)}\circledast\boldsymbol{s(n)}+\boldsymbol{n(n)},
\end{equation}
where $\circledast$ denotes the convolution operation. Convolution in the time domain can be represented using multiplication in the frequency domain. Therefore, the measurement model at frequency bin $k$ can be represented as
\begin{equation}
\boldsymbol{Y(k)}=\boldsymbol{H_c(k)}\boldsymbol{S(k)}+\boldsymbol{H_t(k)}\boldsymbol{S(k)}+\boldsymbol{N(k)},
\end{equation}
where $\boldsymbol{H_c(k)}$ and $\boldsymbol{H_t(k)}$ denote the clutter and target stochastic transfer functions respectively.

\begin{figure*}[htbp!]
    \centering
    \includegraphics[scale=1.5]{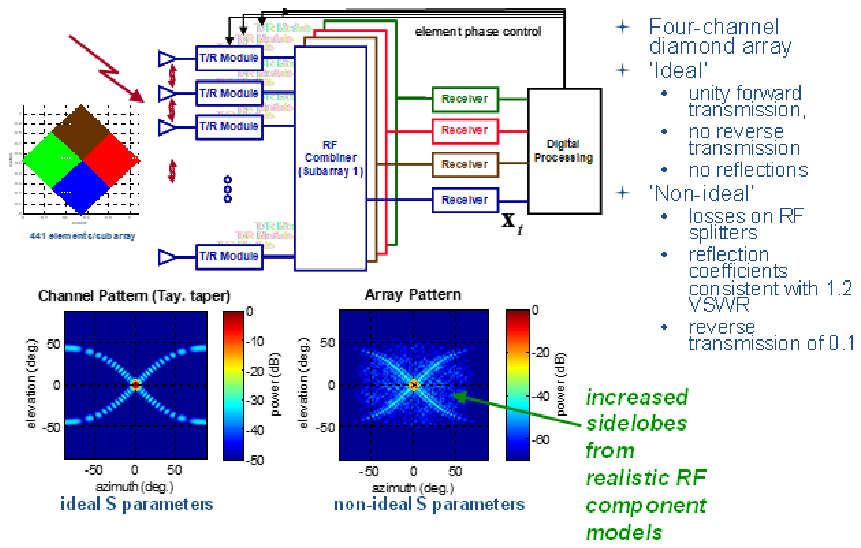}
    \caption{Example of high-fidelity Active Electronically Scanned Array (AESA) that captures many real-world RF imperfections.}
    \label{aesa}
    \end{figure*}

Having zeroed-in on the physics based linear model from the above equation, the natural next question would be what goes into the computation of these impulse responses/transfer functions. For the above model to be accurate, the transfer functions must capture the interaction of a transmitted ideal delta function with every component present in the scene. For example, for the clutter channel, the scene (which is typically several square kilometers in size) has to be broken down into extremely small patches and the impact of each individual patch on the received data has to be modeled. The overall clutter returns are a summation of the returns from each individual clutter patch. In other words, the reflectivity of each individual patch along with the propagation attenuation have to be accurately captured to have a realistic model.

\begin{figure*}[htbp!]
    \centering
    \includegraphics[scale=1.2]{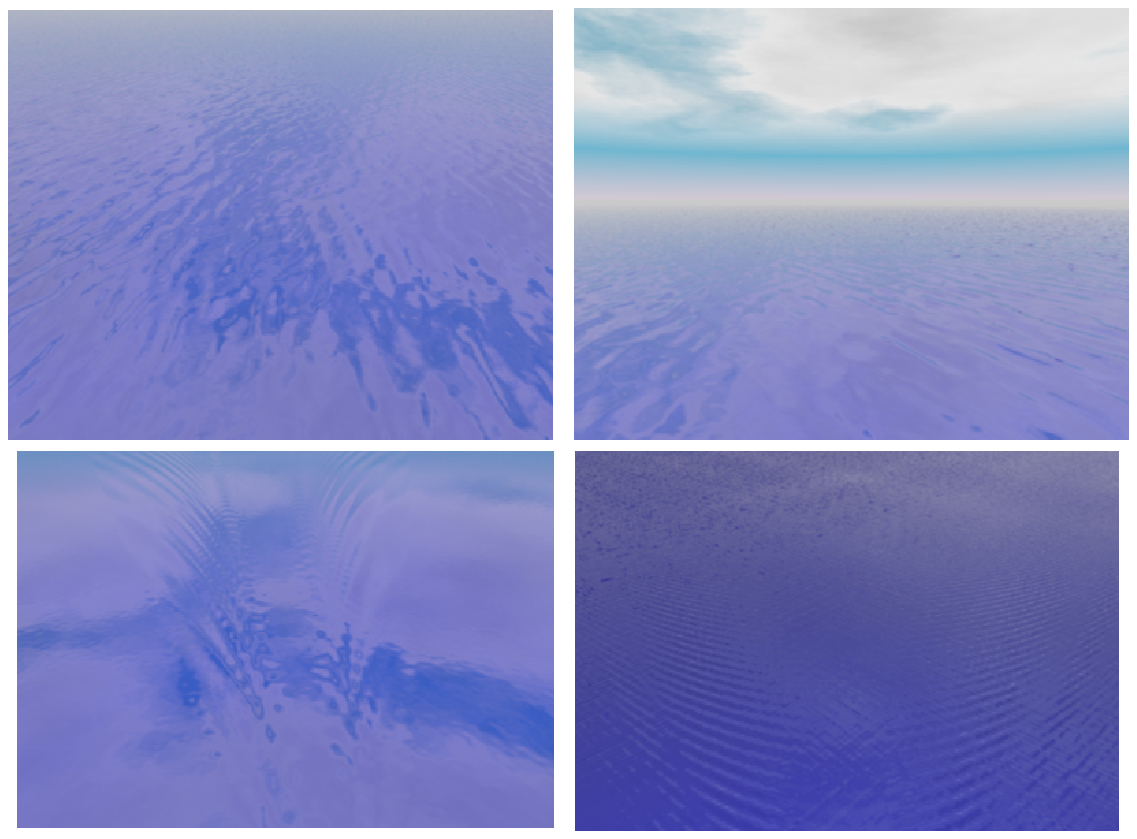}
    \caption{Renderings of ocean surface realizations from PBOSS model. Bottom: Ocean surfaces including Kelvin wakes generated from a moving ship.}
    \label{ocean}
    \end{figure*}

For any given patch, whether there is a line-of-sight (LOS) component based on the transmitter and receiver locations needs to be determined first. If an LOS does indeed exist, then the reflectivity off that patch depends on the tilt angle of the patch, operating frequency band, type of material present in the patch, etc. A sophisticated M\&S tool incorporates all these factors while computing the transfer function. For example, Fig. \ref{scatpol} demonstrates the monostatic scattering polynomials as a function the grazing angle for different types of terrain at X-band. For other frequency bands, the scattering polynomials will be quite different. These scattering polynomials have been extensively studied in literature \cite{billingsley}--\nocite{ulaby}\nocite{Watson1}\cite{Watson2}. In addition to the scattering polynomials demonstrated in Fig. \ref{scatpol}, developing scattering models for ocean surfaces involves additional challenges as a result of moving ship-effects. The Physics-Based Ocean Surface and Scattering (PBOSS) model described in \cite{Watson1}, \cite{Watson2} incorporates both environmental conditions (i.e. atmospheric and oceanographic) and moving ship effects (i.e. Kelvin and near-field/narrow-V wakes) to generate realizations of the ocean surface and the spatially-varying scattering properties of the ocean. The PBOSS model has been incorporated into the high-fidelity M\&S tool RFView \cite{rfview} to provide RF phenomenology characterization of ocean environments. We can leverage this existing modeling capability, incorporating the effects of the surrounding ocean surface and its scattering properties, to adaptively and optimally design waveforms for detection of submarines and ships. An example rendering of the ocean surface model output using a ray tracer \cite{Gonzato}, \cite{Qu} is shown in Fig. \ref{ocean}. Clouds, sky, and fog are included in the rendering for realism.

Additionally, the impact of the propagation medium needs to be implemented along with the scattering model. The channel is highly dependent on the characteristics of the propagation path between the radar and the targets and clutter patches in scene. At higher frequencies such as X-band the propagation is often dominated by line-of-sight and can be approximated by simple models that identify blockages along the propagation path caused by terrain and buildings in the scene and simply apply a larger attenuation to ‘shadow’ regions. When higher fidelity is required or when simulating systems operating a lower frequencies it may be necessary to include more advanced propagation modes such as multipath, diffraction and ducting. Modeling these modes typically involves analysis of the terrain profile between the radar and target (or clutter patch) to determine the most appropriate mode or combination of modes of propagation to employ for predicting the propagation loss. A good example of this type of model is the SEKE \cite{Ayasli} model developed by MIT Lincoln Laboratory. This model includes multiple knife-edge diffraction, spherical earth diffraction, and multipath to predict the site-specific propagation loss along a specified terrain profile typically extracted from a terrain databases such as DTED.

While these types of models are somewhat ad hoc, they are relatively computationally efficient and can provide very realistic results. Effects such as ducting which can dominate propagation in environments with more complex atmospheric conditions typically require more sophisticated and generally more computationally intensive methods such as parabolic wave equation solutions \cite{Dockery}. The Advanced Propagation Model (APM) \cite{Barrios} developed by the SPAWAR Systems Center is an example of a propagation code that includes this type of mode. This APM code allows the atmosphere to be specified with an index of refraction that varies both vertically and horizontally within the plane of propagation between a radar and target/clutter. This allows for accurate simulations of the well-known ducting phenomenon often encountered in maritime and littoral environments.

In addition to high-fidelity environmental modeling, precision modeling of all RF subsystems and components is crucial to again capture many real-world effects. For example, Fig. \ref{aesa} shows the difference (antenna pattern) between a standard aperture model using approximations and idealizations, and one that includes a variety of real-world RF component imperfections. This degree of realism is essential if the simulated data is to serve as a testbed for radar algorithms. After incorporating all these real-world hardware effects, scattering, propagation models and environmental interactions with ground clutter, an advanced M\& S tool needs to compute the raw I \& Q measurements at the radar receivers along with the true EM propagation channel impulse responses. Having described the stochastic transfer function based radar clutter model, we will now present several examples in the next section spanning multiple RF applications.

\begin{figure}[htbp!]
    \centering
    \includegraphics[scale=1]{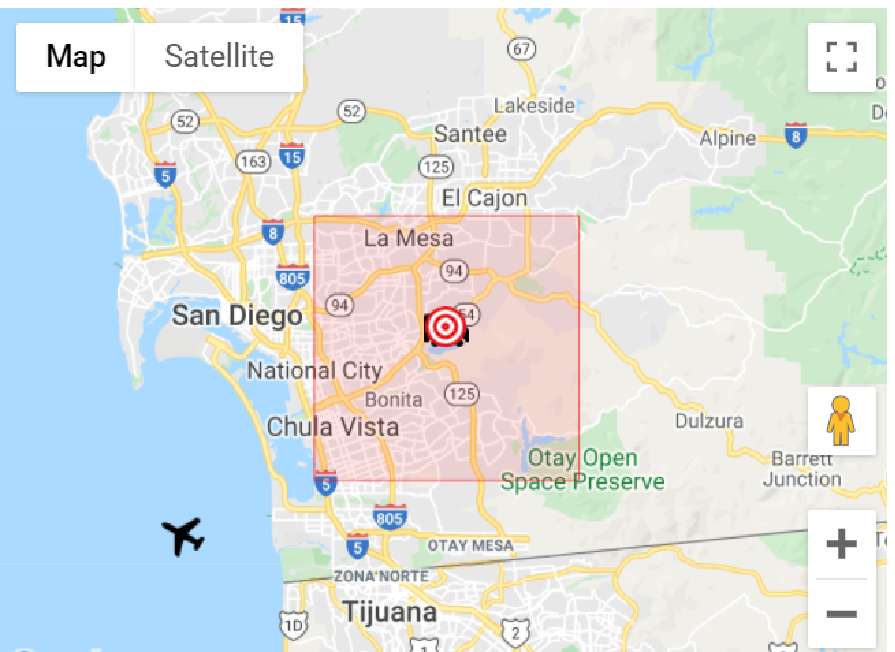}
    \centering
    \caption{Google maps illustration of the simulated monostatic scene with airborne radar and ground target.}
    \label{gmaps}
    \end{figure}

\section{Modeling and Simulation Examples}
All the examples presented in this paper have been generated using high-fidelity RF modeling and simulation tool RFView \cite{rfview} which generates the data using the stochastic transfer function model presented in the previous section.

\subsection{Ground Moving Target Indicator (GMTI) Radar}
We start with a monostatic GMTI radar example. Monostatic radar systems have colocated transmitter and receiver. We consider an airborne X-band (10 $\mathrm{GHz}$) radar flying along the coast of southern California looking at a ground moving target (see Fig. \ref{gmaps}). The radar is moving at an altitude of $1000\mathrm{m}$ with a speed of $125\mathrm{m/s}$. The simulation spans a range swath of $20\mathrm{km}$ with a linear frequency modulated (LFM) waveform of bandwidth $5\mathrm{MHz}$ and $65$ pulses with a pulse repetition frequency (PRF) of $2100 \mathrm{Hz}$. We now look at the different layers that go into the calculation of the impulse responses. Firstly, for each patch in the scene, the presence or absence of an LOS component needs to be computed.

\begin{figure}[htbp!]
    \centering
    \includegraphics[scale=0.2]{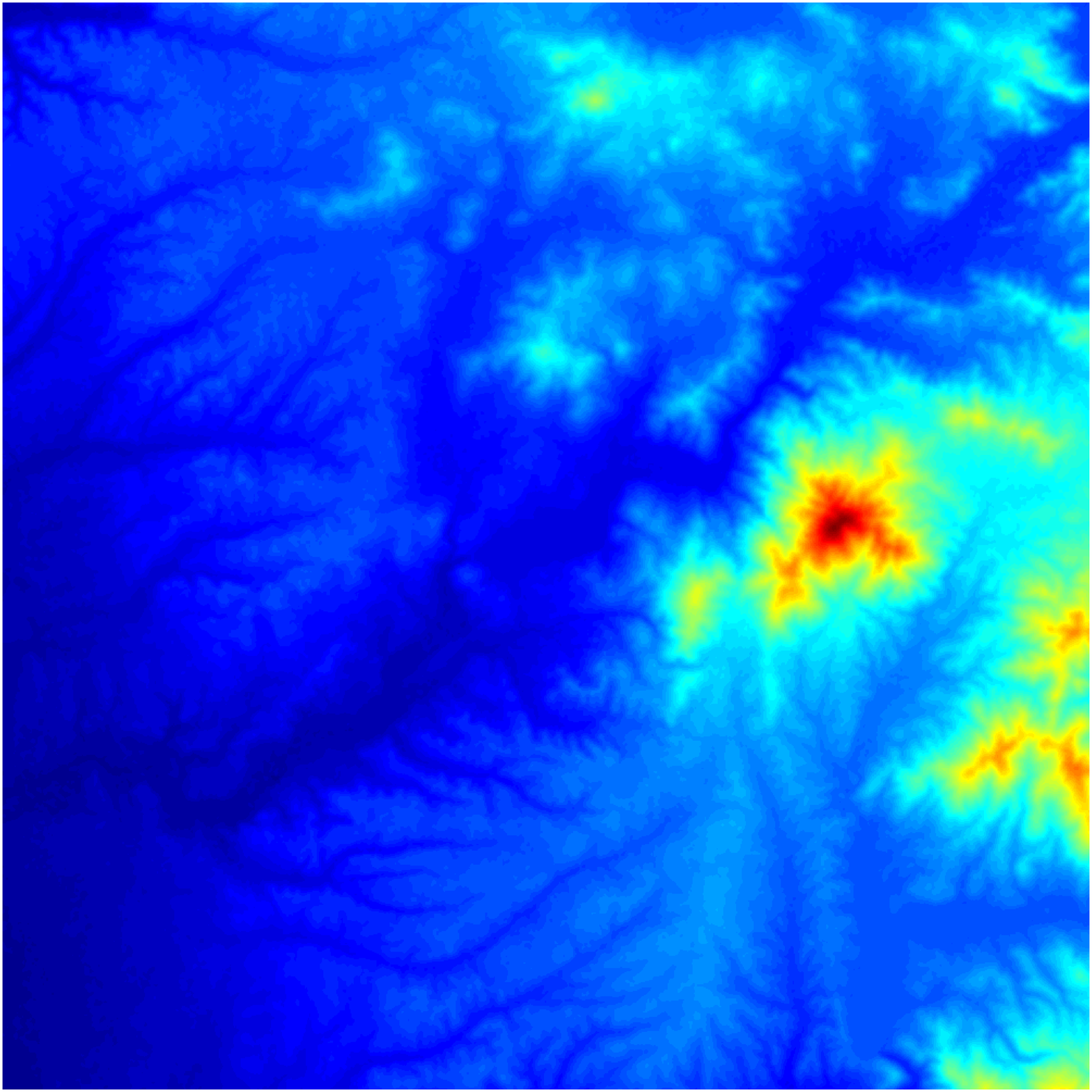}
    \caption{Terrain map of the simulated scene.}
    \label{terrain}
    \end{figure}

\begin{figure}[htbp!]
    \centering
    \includegraphics[scale=0.2]{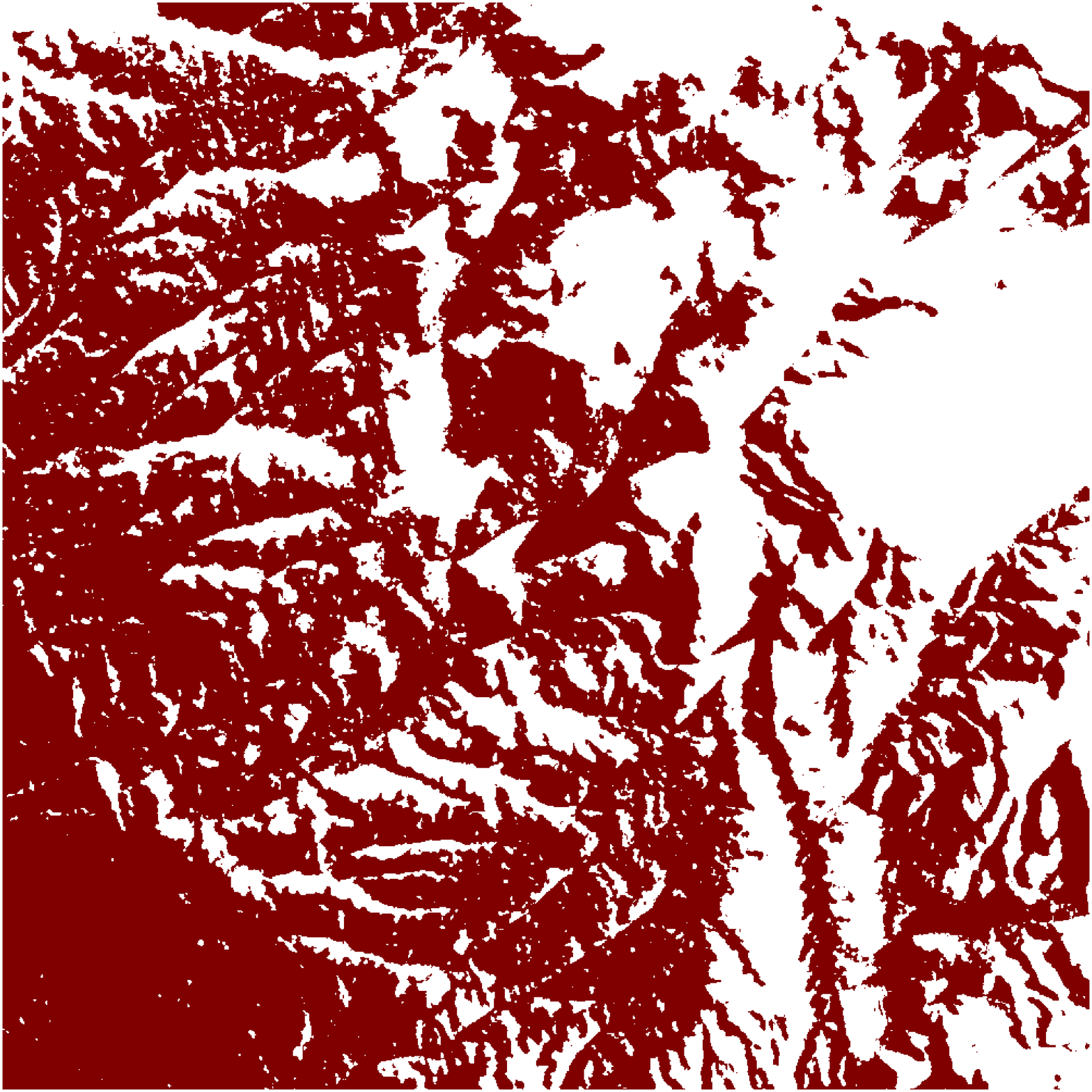}
    \caption{LOS map of the simulated scene.}
    \label{los}
    \end{figure}

Fig. \ref{terrain} shows the terrain map of the simulated scene. We can clearly notice that the scene has mountainous terrain and one huge mountain peak that is denoted by the red region in the terrain map. This information on the terrain is obtained from publicly available terrain databases and they span the entire earth. Given this terrain map, for each patch, the LOS map is demonstrated in Fig. \ref{los}. We can observe that the region behind the mountain peak is shadow region that cannot be penetrated/illuminated by the radar signal. Hence, the shadow regions do not contribute to the clutter returns.

    \begin{figure}[htbp!]
    \centering
    \includegraphics[scale=0.2]{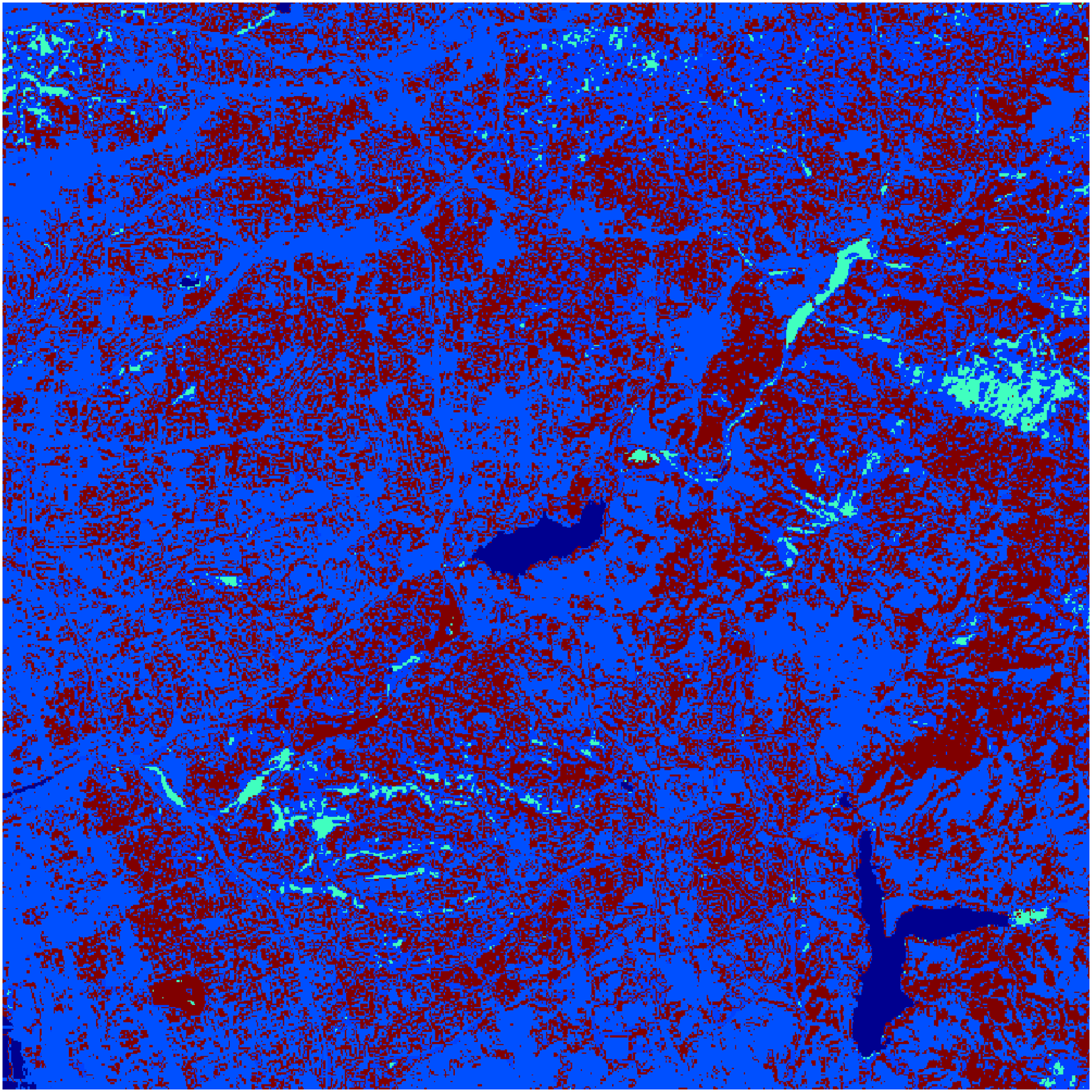}
    \caption{Land cover map of the simulated scene.}
    \label{cover}
    \end{figure}

    \begin{figure}[htbp!]
    \centering
    \includegraphics[scale=0.2]{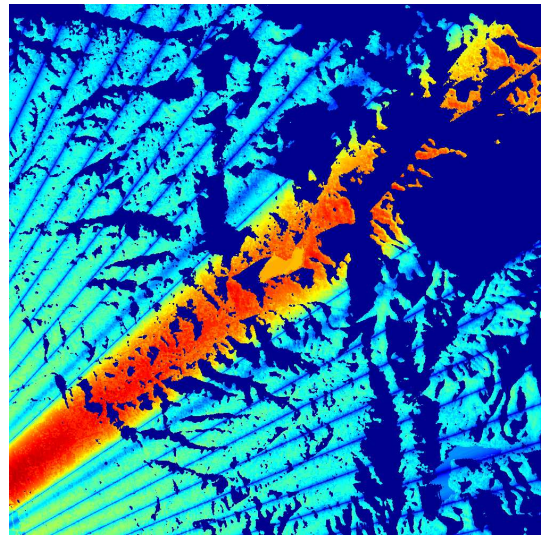}
    \caption{Final clutter map of the simulated scene.}
    \label{cmap}
    \end{figure}
Next, for the patches which indeed have an LOS component, the reflectivity has to be computed using the scattering polynomials described in the previous section. As mentioned earlier, these scattering polynomials vary for different types of terrain. Hence, it is important to use environmental databases that describe the type of terrain present in each clutter patch. Fig. \ref{cover} demonstrates the different types of terrain present in each patch. Each terrain type leads to a unique clutter response and leads to the overall clutter map with reflectivity from each patch demonstrated in Fig. \ref{cmap}. Note that this clutter map also shows the effect of the radar main beam and side lobes and the reflectivity from each patch also depends on the incident energy from the radar beam.

    \begin{figure}[htbp!]
    \centering
    \includegraphics[scale=1.2]{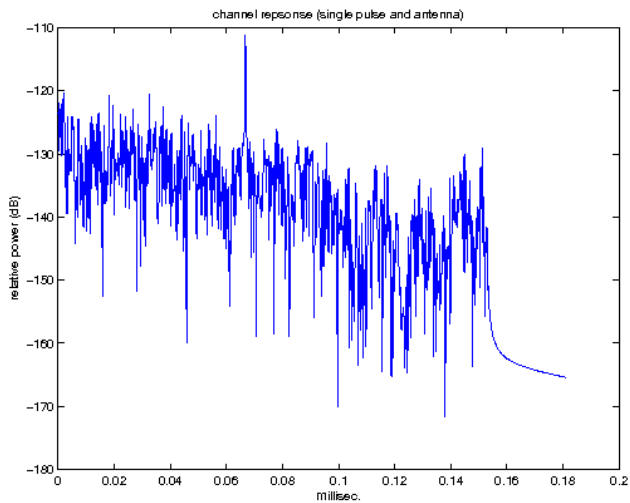}
    \caption{Green's function impulse response of the clutter+target channel for the simulated monostatic GMTI example.}
    \label{imp}
    \end{figure}

Having demonstrated the different components that go into the computation of the RF clutter map for this monostatic GMTI example, we now calculate the Green's function impulse response for the ground clutter and target channels. For the clutter channel, the impulse response is computed as a summation of the responses from each individual clutter patch in the clutter map. Note that along with the reflectivity, each patch also induces a different delay and Doppler component on the incident RF signal. Typically any given scene can contain hundreds of thousands or even millions of patches. However, due to the inherently parallel nature of the computations, the impulse responses can be computed in near real-time using high-performance-computing clusters or GPUs. Recent advances in accelerated computing making it feasible to use these advanced methods for realistic RF clutter simulation.For this example, Fig. \ref{imp} demonstrates the impulse response that is summation of both the clutter and target channels. Note that while the impulse response itself typically has a complex amplitude, we plot the corresponding power in Fig. \ref{imp}. As we can clearly see, the one big peak corresponds to the target and it shows up at the appropriate time instance based on the location of the target. The rest of the impulse response is specific to the clutter scene that has been simulated in this example. Note that the response at any range-bin in the impulse response can be a cumulative effect of responses from multiple clutter patches.

 \begin{figure}[htbp!]
    \centering
    \includegraphics[scale=1.2]{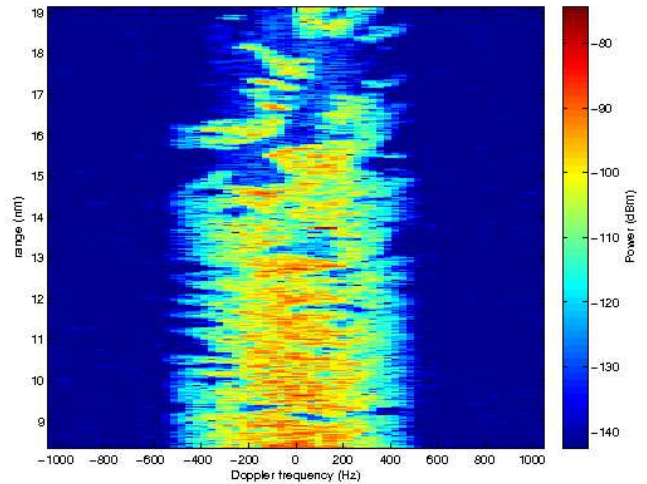}
    \caption{Range-Doppler plot after processing the raw IQ data generated by the simulator.}
    \label{rng_dop}
    \end{figure}

Given the impulse response displayed in Fig. \ref{imp}, we can calculate the raw IQ data at the radar receiver as a convolution of the radar transmit waveform and the impulse response with additive noise. Note that the impulse response is site-specific and accurately captures all the local features present in the simulated scene. Therefore, the IQ data generated using this approach is very realistic compared to approximate statistical methods that have been used for several decades. Simple beamforming and matched filtering of the data generated using the simulator produces the range Doppler plot demonstrated in Fig. \ref{rng_dop}. The patterns of clutter that show up in this plot are again site-specific and if we were to repeat this example at a different location, the generated plots would match the operating environment instead of just using average statistics for several scenes.

\begin{figure}[htbp!]
    \centering
    \includegraphics[scale=1]{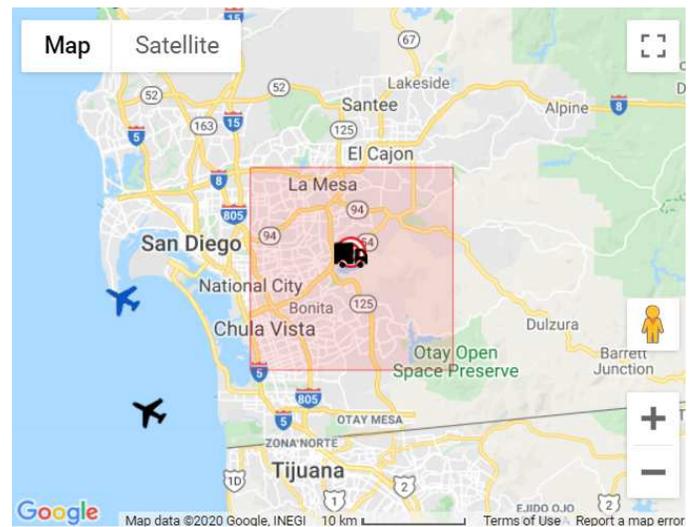}
    \centering
    \caption{Google maps illustration of the simulated bistatic scene with airborne radar transmitter (blue), receiver (black), and ground target.}
    \label{bis_gmap}
    \end{figure}

Next, we present a bistatic GMTI example. In a bistatic scenario, the radar transmitter and receiver are present in different physical locations. As a result of this, the underlying computations for the scattering coefficients of each patch present in the scene is completely different compared to the monostatic case. Even the LOS computations need to take into account the presence of a direct path from both the transmitter and receiver to the simulated patch. Therefore, the shadow regions will also be different. We consider the same scenario and radar parameters as we used for the monostatic example above. However, now the transmitter is moved to a different location as shown with a blue aircraft symbol in Fig. \ref{bis_gmap}. The transmitter is flying at same altitude of $1000\mathrm{m}$ as the receiver.

    \begin{figure}[htbp!]
    \centering
    \includegraphics[scale=1.2]{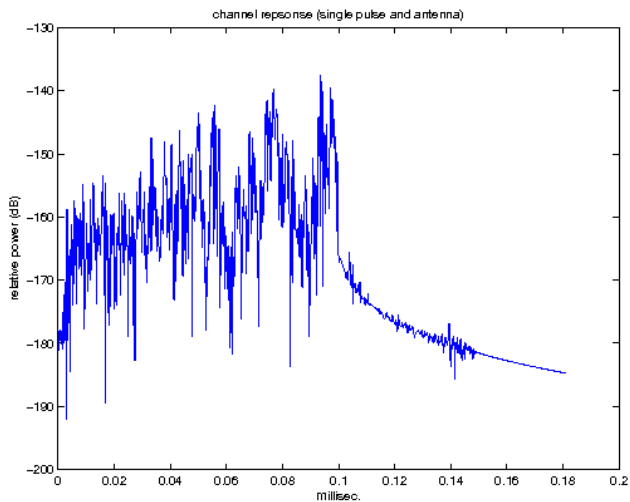}
    \caption{Channel impulse response for the simulated bistatic GMTI example.}
    \label{imp_bi}
    \end{figure}

\begin{figure}[htbp!]
    \centering
    \includegraphics[scale=1.2]{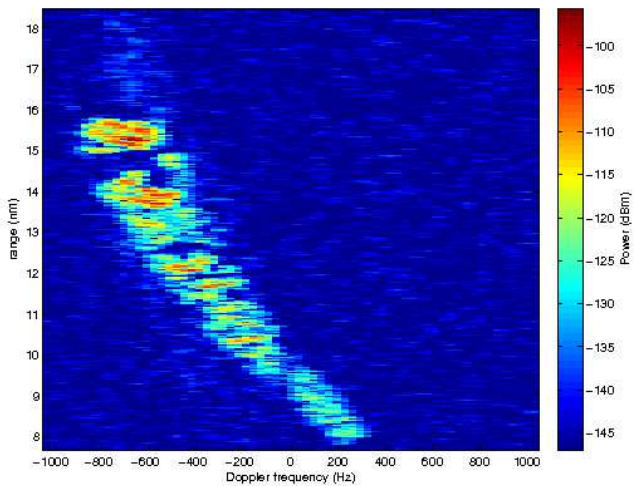}
    \caption{Range-Doppler plot after processing the raw IQ data generated by the simulator for bistatic example.}
    \label{rng_dop_bi}
    \end{figure}

We plot the channel impulse response for this bistatic GMTI example in Fig. \ref{imp_bi} and clearly notice the difference compared to the impulse response for the monostatic example in Fig. \ref{imp}. Similarly, after processing the receiver IQ data, we obtain the range Doppler plot in Fig. \ref{rng_dop_bi}. As expected, this range Doppler plot captures the effects of the bistatic geometry of the simulation. The delays and Doppler frequencies are now a function of the locations of both the airborne transmitter and receiver.

\subsection{Cognitive Fully Adaptive Radar (CoFAR)}
Both the examples in the previous sub-section represent traditional radar systems with fixed transmit functions. We now simulate a more advanced radar system called CoFAR or CR in short. It is very important for and M\&S tool to simulate emerging technologies and systems along with the traditional ones. In fact, emerging technologies and algorithms need the most data for testing and evaluation. Cognitive radar (CR) has emerged as key enabling technology to meet the demands of ever increasingly complex, congested, and contested radio frequency (RF) operating environments \cite{GuerciBook}. While a number of CR architectures have been proposed in recent years, a common thread is the ability to adapt to complex interference/target environments in a manner not possible using traditional adaptive methods.

For example, in conventional space-time adaptive processing (STAP) it is assumed that a sufficient set of Wide Sense Stationary (WSS) training data is available to allow for convergence of the adaptive weights \cite{GuerciBook2}. Though a variety of “robust” or reduced-rank training methods have been proposed over the past $25$ years \cite{GuerciBook2}, there are still many real-world scenarios where even these methods are insufficient. These environments are routinely encountered, for example, in dense urban and/or highly mountainous terrain, and/or in highly contested environments. In contrast, CR uses a plurality of advanced knowledge-aided (KA) and artificial intelligence (AI) methods to adapt in a far more sophisticated and effective manner.

For example, in urban terrain, a KA CR would have access to a detailed terrain/building map and real-time ray-tracing tools in order to adapt with extreme precision to targets anywhere in the scene, even those behind buildings (see \cite{Fertig} and \cite{WatsonBook} for recent work in this area). In cognitive fully adaptive radar (CoFAR) \cite{GuerciBook}, the presence of a fully adaptive transmitter allows for active multichannel probing to support advanced signal-dependent channel estimation. All CR architectures have some form of a Sense-Learn-Adapt (SLA) decision process. The latest CR architectures differ mainly in the ways in which each of these steps is performed.

\begin{figure*}[htbp!]
    \centering
    \includegraphics[height=8cm]{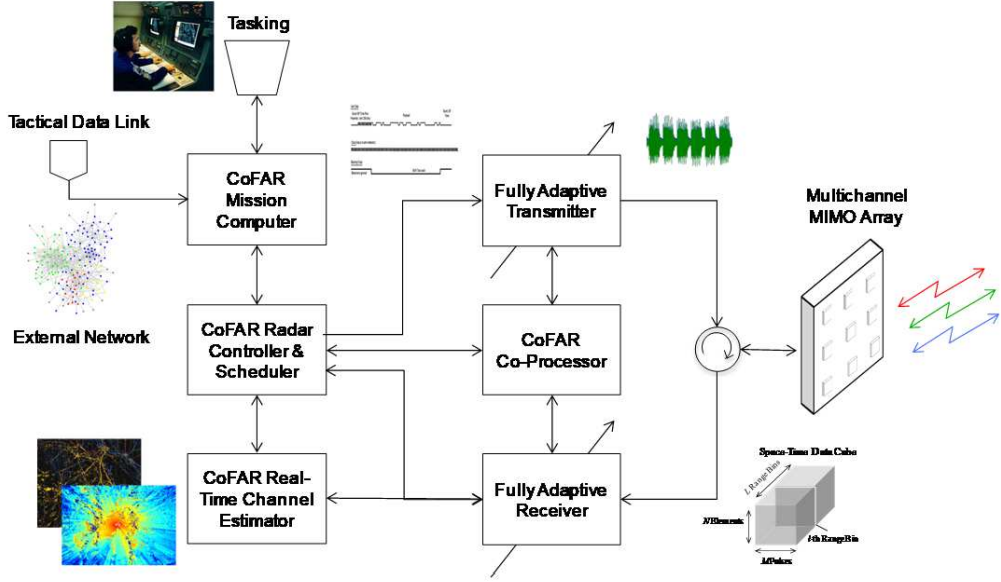}
    \caption{Major elements of a cognitive fully-adaptive radar (CoFAR)}
    \label{cr}
    \end{figure*}

Fig. \ref{cr} shows, at a high level, some of the major elements of a CoFAR system. Key processing elements include:
\begin{itemize}
  \item CoFAR Radar Controller and Scheduler: Performs optimal real-time resource allocation and radar scheduling. It receives mission objectives and has access to all requisite knowledge-bases and compute resources to effectively enable optimal decisioning.
  \item CoFAR Real-Time Channel Estimator: Performs advanced multidimensional channel estimation using a plurality of methods including KA processing, real-time ray-tracing, active MIMO probing, and/or machine learning techniques.
  \item CoFAR Co-Processor: Performs extremely low-latency adaption (potentially intra-pulse). Mostly applicable in advanced electronic warfare applications.
  \item Fully Adaptive Receiver: Features the usual adaptive receiver capabilities such as adaptive beamforming, pulse compression, etc.
  \item Fully Adaptive Transmitter: A relatively new feature to radar front-end. Extremely useful for pro-active channel probing and support of advanced adaptive waveforms.
\end{itemize}

In many respects, a key goal of all the above is channel estimation. “As goes channel knowledge, so goes performance”. In the CoFAR context, the channel consists of clutter (terrain, unwanted background targets), targets, atmospheric, meteorological effects, and intentional, and/or unintentional RFI. To capture real-world environmental effects, and to present the CR with a meaningfully challenging simulation environment, clutter often presents the greatest challenge. The Green's function impulse response method that we described in the previous section exactly addresses this issue and provides an accurate site-specific testbed to evaluate these advanced CoFAR techniques. In this paper, we present one such example that involved an advanced CoFAR system that optimally adapts its transmit waveform to match the operating environment to achieve optimal radar performance. This is in contrast to traditional radar systems that transmit a fixed waveform.

Given the measurement model in the previous section, the goal of CR waveform design is to find the optimal waveform $\boldsymbol{S}$ (stacked into a vector) such that it maximizes the signal-to-clutter-plus-noise-ratio (SCNR) subject to energy constraint $\boldsymbol{S}^{H}\boldsymbol{S}=1$
\begin{equation}
\mathrm{SCNR}=\frac{E\left\{{\|\boldsymbol{H_tS}\|}^2\right\}}{E\left\{{\|\boldsymbol{H_cS+N}\|}^2\right\}},
\end{equation}
where $\boldsymbol{H_t}$ and $\boldsymbol{H_c}$ denote the target and clutter channel stochastic transfer functions and $\boldsymbol{N}$ denotes the additive noise. Also, $E\left\{.\right\}$ denotes the expectation operator. Note that the transfer functions still have a random component that can be induced by intrinsic clutter motion and other factors which is why we use the expectation operator. The solution to this optimization problem can be easily shown to satisfy the following generalized eigenvector formulation
\begin{equation}
\lambda \left(E\left\{\boldsymbol{H_c}^H\boldsymbol{H_c}\right\}+\sigma^2 \boldsymbol{I}\right)\boldsymbol{S}=E\left\{\boldsymbol{H_t}^H\boldsymbol{H_t}\right\}\boldsymbol{S},
\end{equation}
where $\sigma^2$ denotes the additive noise variance. Since the matrix $E\left\{\boldsymbol{H_c}^H\boldsymbol{H_c}\right\}+\sigma^2 \boldsymbol{I}$ is always invertible, we can further write down the optimal waveform as the eigenvector of the following matrix
\begin{equation}
\left(E\left\{\boldsymbol{H_c}^H\boldsymbol{H_c}\right\}+\sigma^2 \boldsymbol{I}\right)^{-1}E\left\{\boldsymbol{H_t}^H\boldsymbol{H_t}\right\}\boldsymbol{S}=\lambda\boldsymbol{S}.
\end{equation}

As we can see from the measurement model described in Fig. \ref{stf_model}, the radar clutter and target channel impulse responses (or transfer functions) are independent of the transmit waveform itself even though the IQ data at the receiver is signal-dependent clutter. This approach to modeling makes it feasible to generate simulated data for any arbitrary waveform that has been designed by CR. The new choice of waveform (one example of optimal waveform design is described above) will be convolved with the appropriate channel impulse responses. This ability to simulate realistic radar data for rapidly adapting radar waveforms makes this approach to M\&S a good match to test different CR algorithms without having to resort to expensive measurement campaigns which are further limited by the number of algorithms that can be tested in a single data collect mission. The channel impulse responses corresponding to different locations on earth can be simulated to test the generality of the CR techniques. Further, multiple CPIs of data can be generated to simulate the changing dynamics of these channel impulse responses and their impact on the CR performance.

\begin{figure*}[htbp!]
    \centering
    \includegraphics[height=6cm]{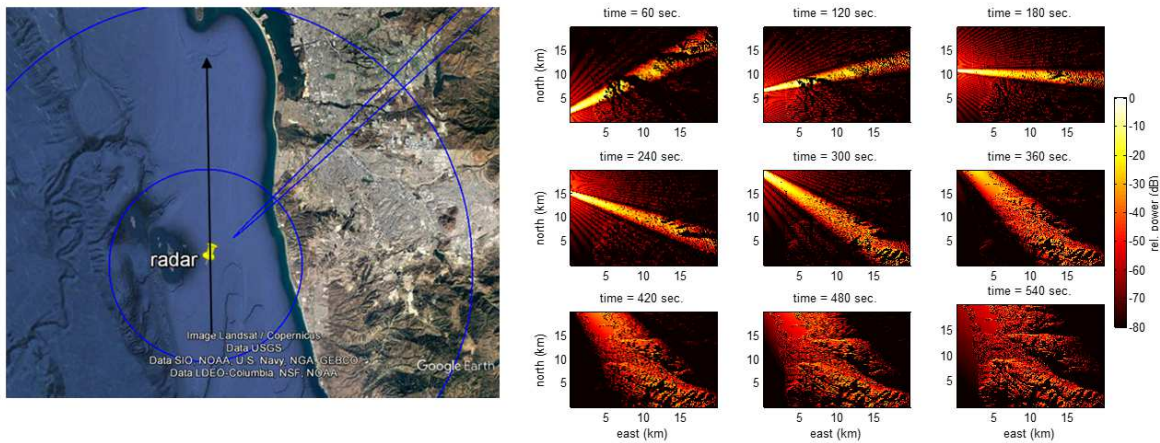}
    \caption{X-band site-specific airborne GMTI radar scenario off the coast of southern California. Left: Scenario location and geometry. Right: Radar beam pointing positions at different portions of the flight.}
    \label{cr1}
    \end{figure*}

    \begin{figure*}[htbp!]
    \centering
    \includegraphics[height=6cm]{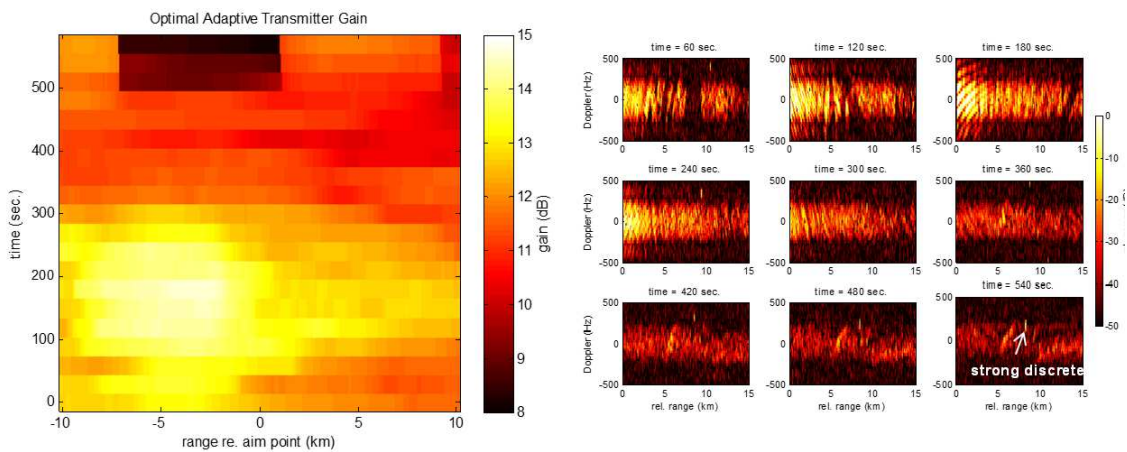}
    \caption{Left: Optimal maximum gain (dB) using adaptive waveforms as a function of time and range of interest. Note in general maximum gain is achieved in regions with the strongest heterogenous clutter (as expected). Right: Corresponding range-Doppler plots. Note that there is no gain when the clutter is weak with the presence of a single discrete (again to be expected).}
    \label{cr2}
    \end{figure*}

Fig. \ref{cr1} shows an airborne X-band radar surveillance scenario of a northbound offshore aircraft flying off the coast of southern California. The region has significant heterogenous terrain features and thus presents an interesting real-world clutter challenge. In the presence of flat terrain with no clutter discretes, the spectrum will be flat and as a result there wouldn't be any advantage as a result of adapting the transmit waveform. The heterogenous terrain in this example ensures strong CoFAR performance potential. Shown in Fig. \ref{cr2} is the theoretical performance gain (tight bound) using the optimal pulse shape prescribed above (max gain = ratio of max to min eigenvalue in dB). As expected, maximum potential gain is achieved in those regions with the strongest heterogenous clutter since this produces significant eigenvalue spread. Also as expected, there is no gain in areas of weak clutter where only a single large discrete (impulse) is present since this yields a flat eigenspectrum. Note however that these results assume the optimizer has access to the true channel transfer functions. However, in reality, these channel impulse responses and their corresponding transfer functions are not known ahead of time and they need to be estimated from the measured data. This may involve sending out strong probing signals to enhances the accuracy of channel estimation. This topic is beyond the scope of this paper and more details on channel estimation can be found in \cite{GoginChanEst2019}--\nocite{GuerciChanEst2019}\nocite{Saeid}\cite{BosungIET}.

\subsection{Multiple Input Multiple Output (MIMO) Radar}
MIMO radar involves multiple transmitters and receivers which can either be co-located or spatially distributed. Further, each transmitter in a MIMO radar has the flexibility to transmit its own unique waveform to achieve best possible performance. The Green's function M\&S approach described in this paper can be readily applied to simulate any MIMO system because a MIMO can be easily broken down into multiple constituent bistatic pairs. Additionally, in each of these individual bistatic channels, the measurement model in Fig. \ref{stf_model} ensures that the waveform is independent of the channel impulse response. Hence, the simulator computes the channels for each bistatic component of the MIMO radar and those channel impulse responses can be convolved with different waveforms corresponding to the different MIMO transmitters.

\begin{figure*}[htbp!]
    \centering
    \includegraphics[height=6cm]{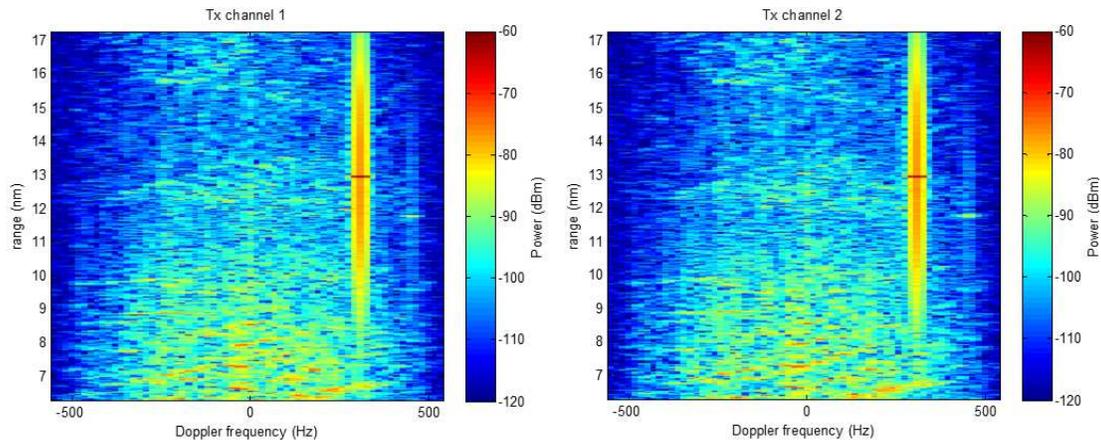}
    \caption{Range Doppler plots for both transmit channels when transmitting LFM waveforms}
    \label{lfm}
    \end{figure*}

\begin{figure*}[htbp!]
    \centering
    \includegraphics[height=6cm]{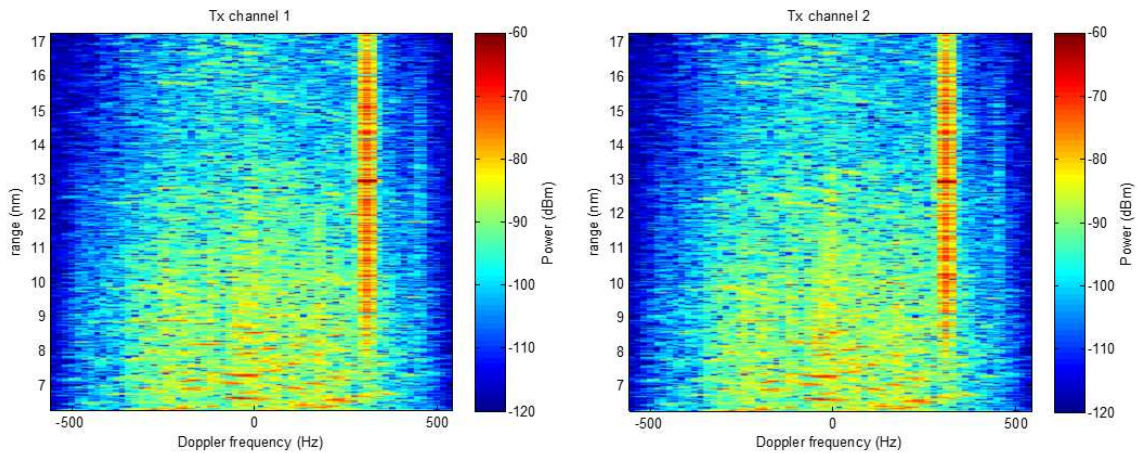}
    \caption{Range Doppler plots for both transmit channels when transmitting random phase coded waveforms}
    \label{pn}
    \end{figure*}

The choice of waveform from the different transmitters is critical to the performance of any MIMO radar. In that context, it is also important to assess the impact of transmitting non-orthogonal waveforms as the signals from multiple transmitters cannot be perfectly separated at the receiver. Such an analysis can be readily made using the M\&S approach presented in this paper. For example, we consider two choices of waveforms for a $2$ transmitter MIMO radar. In the first example, the two transmitters send out LFM waveforms. The first transmitter sends and up-chirp while the second transmitter sends a down-chirp. In the second example, both transmitters send out random phase coded waveforms. As we can see from Figs. \ref{lfm} and \ref{pn}, the waveform dependent features are captured by the M\&S tool because the raw data fed to the processing algorithm is a convolution of the waveforms from each transmitter and the corresponding impulse response of the channel.

\subsection{Ocean Clutter Simulations}
\begin{figure}[htbp!]
    \centering
    \includegraphics[height=6cm]{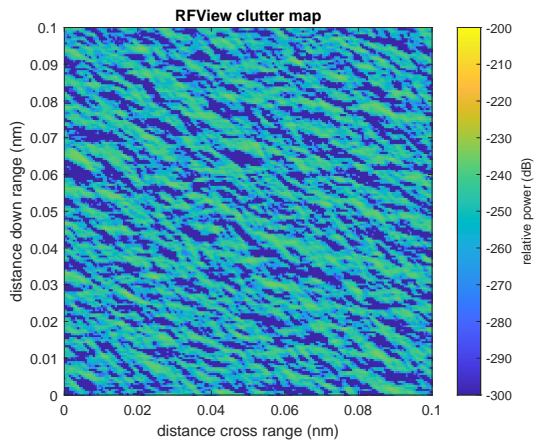}
    \caption{Clutter map of the ocean surface}
    \label{ocean_surface}
    \end{figure}

In Fig. \ref{ocean}, we have shown realistic renderings of ocean surfaces that form the baseline for ocean clutter simulations. However, in a realistic simulation, the surface is extremely dynamic compared to ground terrain simulations. This dynamism can be caused by objects moving on the surface causing wakes and also by the waves caused by the wind speed and direction. In other words, the dynamic nature of the surface itself introduces a Doppler shift on the clutter returns. In Fig. \ref{ocean_surface}, we plot the clutter map corresponding to the ocean surface for a given snapshot. We observe that all the shadow effect caused by the uneven nature of the waves are captured in this simulation. As mentioned above, this surface changes from this snap shot to the next. Therefore, for every pulse, we update the clutter map and then compute the Green's function impulse response for the clutter channel using the updated clutter map for each pulse. This fits nicely into the stochastic transfer function approach to M\&S presented in this paper.

\begin{figure}[htbp!]
    \centering
    \includegraphics[height=6cm]{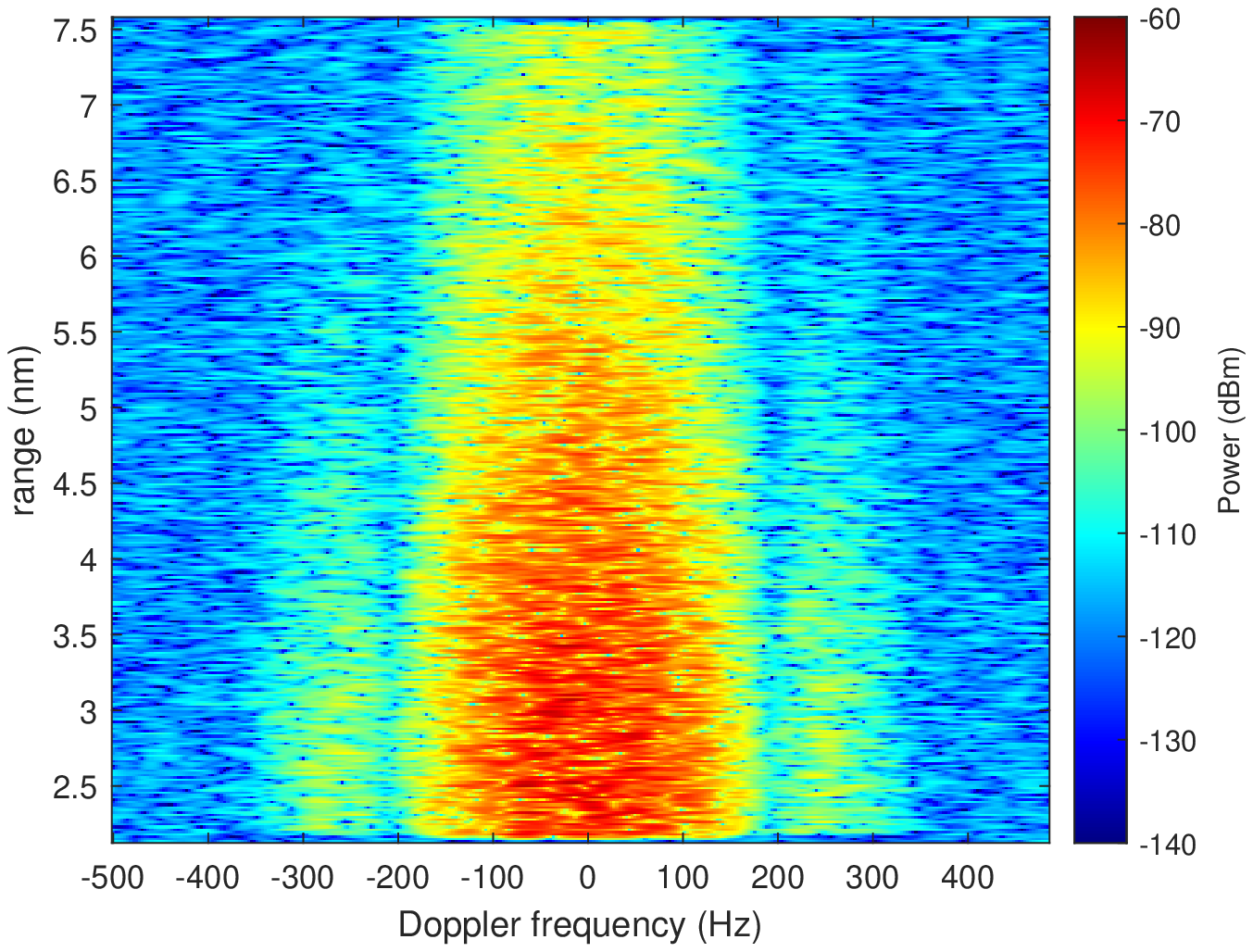}
    \caption{Range Doppler plot from a moving ocean surface with a wind speed of $5 \mathrm{m/s}$}
    \label{moving_ocean_1}
    \end{figure}

 In Fig. \ref{moving_ocean_1}, we generate the Green's function impulse responses (and the data) corresponding to the moving ocean surface for several pulses and then plot the range Doppler map after basic beamforming and matched filtering. The platform was moving at a speed of $30 \mathrm{m/s}$ and the wind in the ocean scene was blowing at $5\mathrm{m/s}$. Similarly, in Fig. \ref{moving_ocean_1}, we produce the same plot but with a wind speed of $30\mathrm{m/s}$. The platform speed is fixed to the same value even for the second simulation. We observe from these plots that the Doppler spread is much significant when the wind speed is larger even though the speed of platform motion is the same. This realistic effect is captured by using dynamic impulse responses (from the M\&S techniques presented in this paper) for each pulse in the simulation.

                \begin{figure}[htbp!]
    \centering
    \includegraphics[height=6cm]{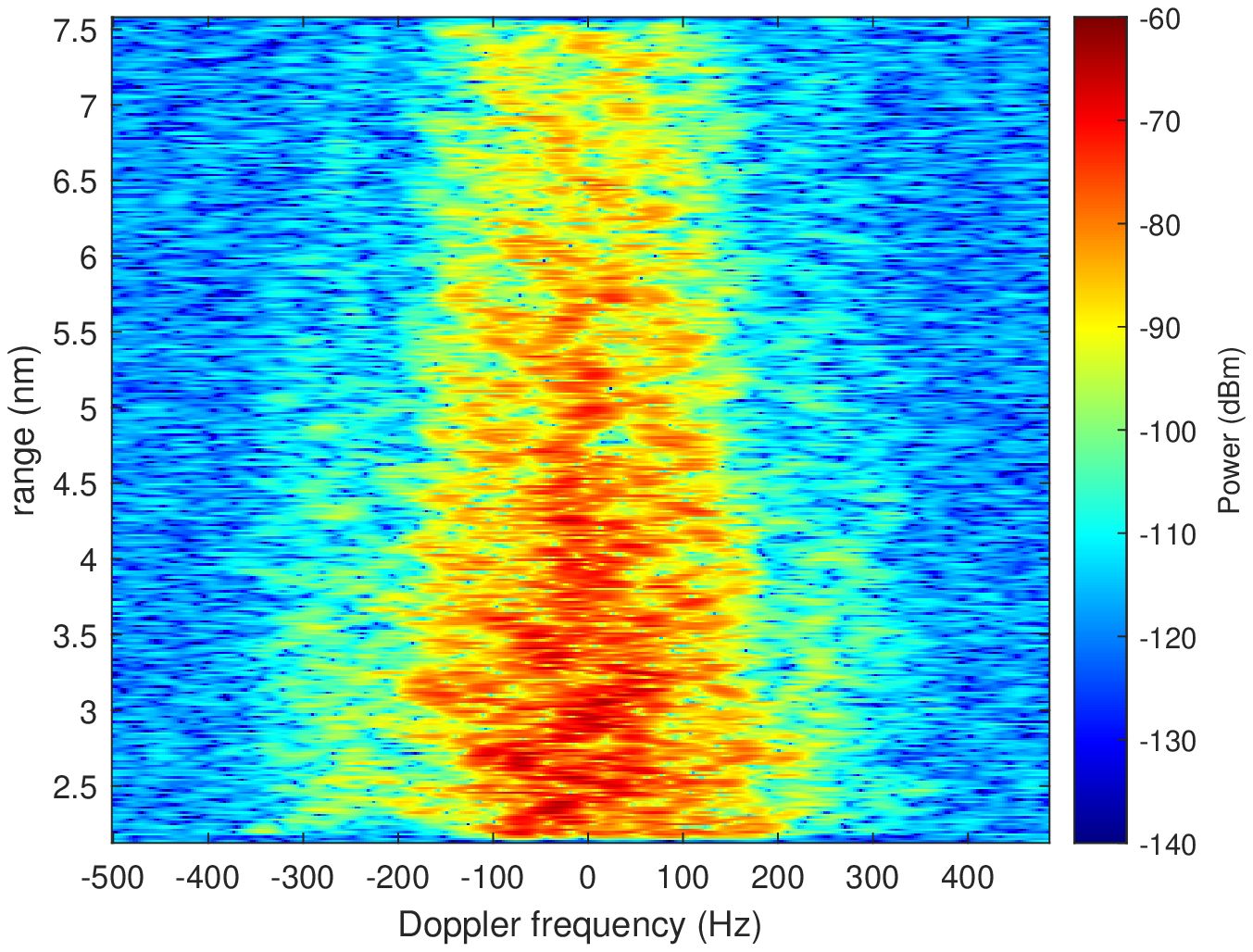}
    \caption{Range Doppler plot from a moving ocean surface with a wind speed of $30 \mathrm{m/s}$}
    \label{moving_ocean_2}
    \end{figure}

                  \begin{figure}[htbp!]
    \centering
    \includegraphics[height=6cm]{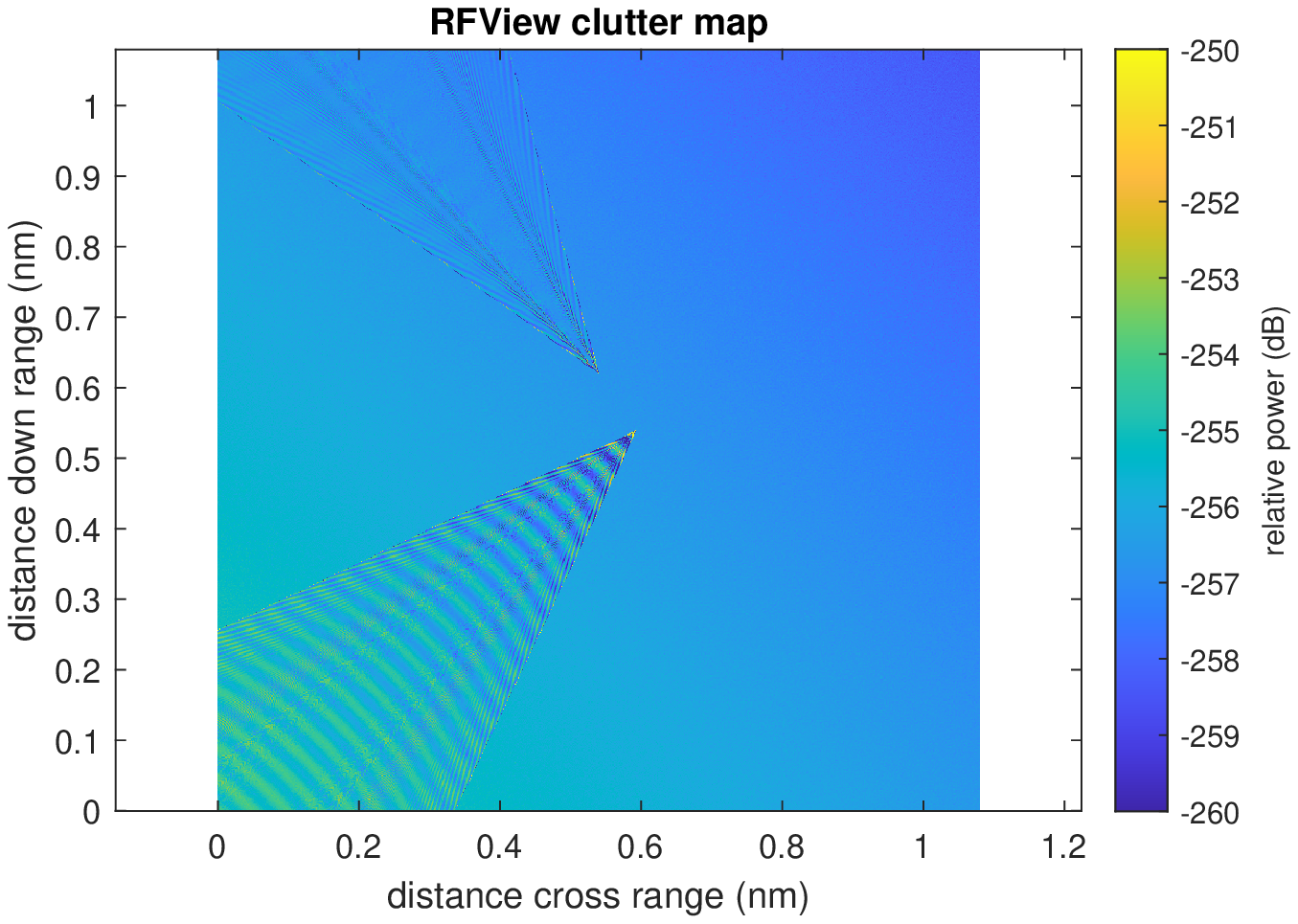}
    \caption{Simulated ocean surface with wakes from multiple ships}
    \label{wakes}
    \end{figure}

As mentioned earlier, along with waves, ocean surfaces also have fluctuations cause by moving objects such as ships in the form of wakes. The shapes of the wakes vary a lot depending on the size of the ship, direction it is moving, etc. In Fig. \ref{wakes}, we plot the ocean surface with two ship wakes on it. Once the ocean surfaces with wakes are generated, we can use the same theory as above to generate the stochastic transfer functions as well as the corresponding I \& Q data.

\subsection{Synthetic Aperture Radar (SAR)}
Synthetic Aperture Radar is typically an airborne or spaceborne radar that generates high-resolution images of a scene or targets by synthesizing a large antenna aperture using a small real antenna aperture of a moving radar. It assumes that the target is stationary and moves the radar position from one position to the next during the data collection time. In SAR, wide bandwidth transmitted signals provide high resolution in range and cross-range/azimuth resolution is obtained by the target’s electromagnetic scattering collected at different aspect angles of a moving radar. Most common operated SAR modes are the strip-map SAR and spotlight SAR, and they can be simulated using the M\&S techniques described above.

\begin{figure}[htbp!]
    \centering
    \includegraphics[scale=0.75]{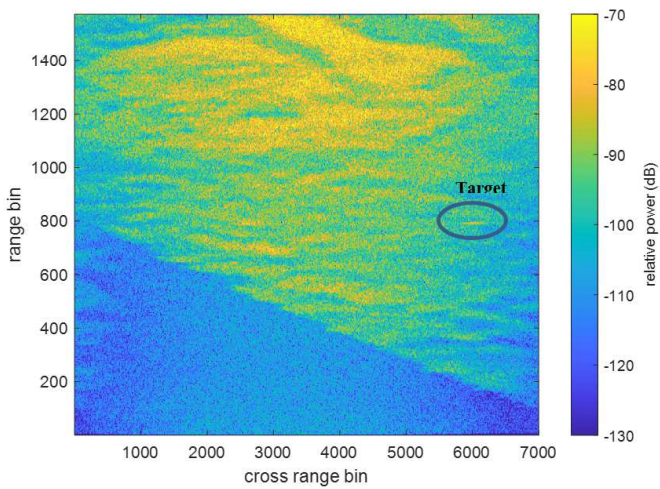}
    \caption{Clutter map of strip-map SAR mode over the coast of San Diego.}
    \label{sar1}
    \end{figure}

In the strip-map SAR mode, the simulator generates wide area maps of the terrain while moving the platform with a fixed pointing direction of the radar antenna.  Fig. \ref{sar1} depicts the strip-map SAR mode clutter map of an airborne X-band radar flying over the San Diego coast. The clutter map was generated by employing the basic range-Doppler processing techniques after motion compensation was conducted using the known platform locations.

\begin{figure}[htbp!]
    \centering
    \includegraphics[scale=0.6]{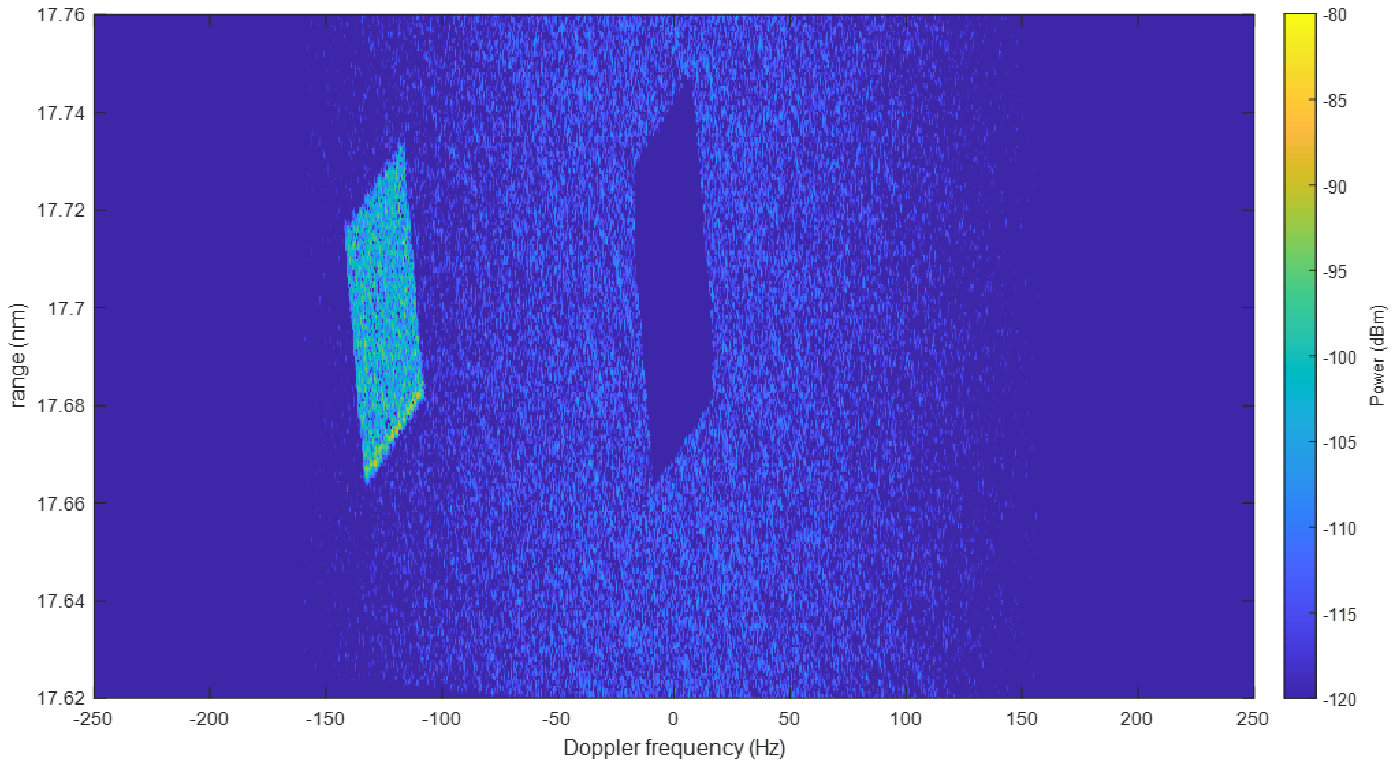}
    \caption{Spotlight SAR mode Example.}
    \label{sar2}
    \end{figure}

In the spotlight SAR, the simulator is flying the platform with narrower beamwidth radar antenna focusing on a fixed location. Fig. \ref{sar2} depicts an example of a range-Doppler spotlight SAR image of an X-band radar flying over the ocean with an iceberg which is moving relative to the background ocean. Similar to the example in Fig. \ref{sar1}, the image in this example was generated from the basic range-doppler processing techniques and motion compensation using the known platform locations. Table 1 provides the radar parameters used in the simulations of the strip-map and spotlight SAR examples.

\begin{table}[!htb]
\captionsetup{size=footnotesize}
\caption{SAR Simulation Parameters} \label{tab:sar}
\setlength\tabcolsep{0pt} 
\footnotesize\centering

\begin{tabular}{ | c | c | c | }
\hline
Radar Parameter & Strip-map SAR & Spotlight SAR \\
\hline
Frequency & 9.6 GHz & 10 GHz \\
\hline
Collection time & 20 seconds & 5 seconds \\
\hline
Bandwidth & 18.75MHz & 150 MHz \\
\hline
Pulse Repetition Frequency (PRF) & 400 Hz & 1000 Hz \\
\hline
Signal Power & 1000 Watts & 1000 Watts \\
\hline
Range Swath & 4 km & 300 m \\
\hline
Range & 47 km & 65 km \\
\hline
Number of Channels & 1 & 3 \\
\hline
\end{tabular}
\end{table}

\subsection{Inverse Synthetic Aperture Radar (ISAR)}
\begin{figure*}[htbp!]
    \centering
    \includegraphics[height=6cm]{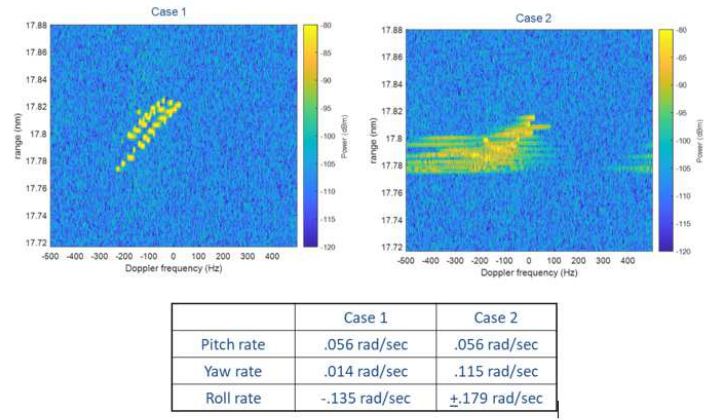}
    \caption{Range-Doppler images of a ship in motion using ISAR mode. Left: Case 1, Right: Case 2}
    \label{isar}
    \end{figure*}

ISAR is typically employed to image targets that are in motions, such as ships, moving ground moving vehicles, aircraft, etc. In ISAR, the synthetic aperture is generated by the motions of the targets relative to the radar and the radar can be stationary on a turret or an airborne platform. The ISAR geometry of a stationary radar and a rotating target is similar to the spotlight SAR geometry of a stationary target and a radar moving in a circular path. However, unlike SAR modes where synthetic aperture is known, synthetic aperture is unknown in ISAR mode due to unknown target’s motion; thus, it is much more difficult forming range-Doppler ISAR images due to unknown motions. To that extent, Fig. \ref{isar} presents the generic range-Doppler images of two ISAR data sets of a ship with movements in the ocean generated using the M\&S techniques presented in this paper. The figure on the left hand side is a ship with some translational and rotational motions where the shape of the ship is still noticeable while the ship on the right hand side figure is unrecognizable due to the fast movements of the ship during the collection time. For the ISAR simulations, the bandwidth is defined as $60 \mathrm{MHz}$ and the collection time is $1$ second while the remaining radar parameters have the same value as the spotlight SAR example above.

\subsection{Real-Time Electromagnetic Environmental Simulator (RTEMES)}
\begin{figure*}[htbp!]
    \centering
    \includegraphics[height=6cm]{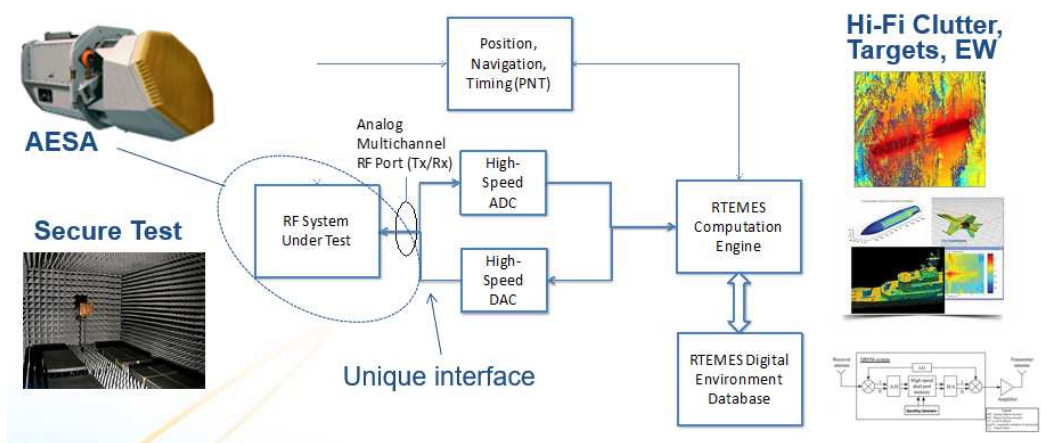}
    \caption{Real-Time Electromagnetic Environmental Simulator (RTEMES) Hardware-in-the-loop (HWIL) implementation.}
    \label{rtemes}
    \end{figure*}

Finally, along with all the different RF modes that we demonstrated in this section, another significant impact of this stochastic transfer function model for RF channels is the ability to easily implement these channels on hardware. Hardware implementation of the channels opens the door for realistic Hardware-in-the-loop (HWIL) testing  (see Fig. \ref{rtemes}) of new radar equipment at a fraction of the cost needed to test using real flight campaigns. By accurately calculating the channel impulse responses for different locations/scenarios and coding them onto FPGA circuits, the radar equipment can be made to believe it is flying over any location on earth. Convolution of the radar transmit signals with the channel impulse responses is performed on the FPGA in real time. This technology has been successfully demonstrated by the RTEMES system in \cite{rtemes}. Along with avoiding expensive flight testing, this new HWIL technology allows us to test radar systems as if they were flying in locations where flight tests would not be possible otherwise. Further, this testing can be performed without leaving the laboratory.

\section{Cognitive Fully Adaptive Radar Challenge Dataset}
In the previous section, we demonstrated various radar applications using the novel stochastic transfer function modeling approach. To conclude this paper, we present a cognitive fully adaptive radar (CoFAR) challenge dataset that was generated using this new modeling approach. High-fidelity, physics-based, site-specific modeling and simulation software RFView® was used to generate the Green’s functions and the corresponding simulated measurements for this dataset. The main purpose of the dataset described here is to provide radar researchers with a common dataset to benchmark their results and compare with existing algorithms.  Along with ground clutter induced by the terrain, we have also included few clutter discretes in the form of buildings. This dataset can be used to test radar detection and estimation algorithms along with CoFAR concepts for radar waveform design. The data was generated using the signal model presented in Section III.

Given this signal model, CoFAR research can be broadly classified into two tasks
\begin{itemize}
  \item Using known channel impulse responses, designing optimal radar transmit waveforms to maximize target detection and estimation performance
  \item Estimating the channel impulse responses from measured data
\end{itemize}
	
Both these tasks are equally important for practical CoFAR systems. Ultimately, in a CoFAR system, allocating resources between these two tasks is a tradeoff. One would like to allocate resources to channel estimation to obtain accurate estimates of the channel impulse responses while not compromising too much on the primary radar objective of target detection and estimation. Some basic ideas for optimal waveform design using impulse-response-based modeling have been discussed in the previous section whereas approaches for channel impulse response estimation have been discussed in \cite{GoginChanEst2019}, \cite{GuerciChanEst2019}. It is our hope that the CoFAR community can develop more advanced waveform design algorithms, further incorporating practical constraints on the waveforms. Our dataset can be used to test the performance of these algorithms for pulse-to-pulse waveform design as well as CPI-to-CPI waveform design. The dataset presented here can be used to perform both the waveform design and channel impulse response estimation tasks.
This dataset contains data cubes as well as the corresponding true channel impulse responses for two scenarios, each of which is described in this section.

This challenge dataset was initially distributed at the inaugural High Fidelity RF Modeling and Simulation Workshop in August $2020$ \cite{workshop}. The dataset can be accessed/downloaded by all readers by creating a free trial account at \cite{rfview}.

\subsection{Scenario 1}
The first scenario is supposed to be a beginner dataset with few targets, ground clutter, and couple of clutter discretes (like buildings). This scenario involves an airborne monostatic radar flying over the Pacific Ocean near the coast of San Diego looking down for ground moving targets. The data spans several coherent processing intervals as the platform is moving with constant velocity along the coastline.

\begin{figure}[htbp!]
    \centering
    \includegraphics[scale=0.8]{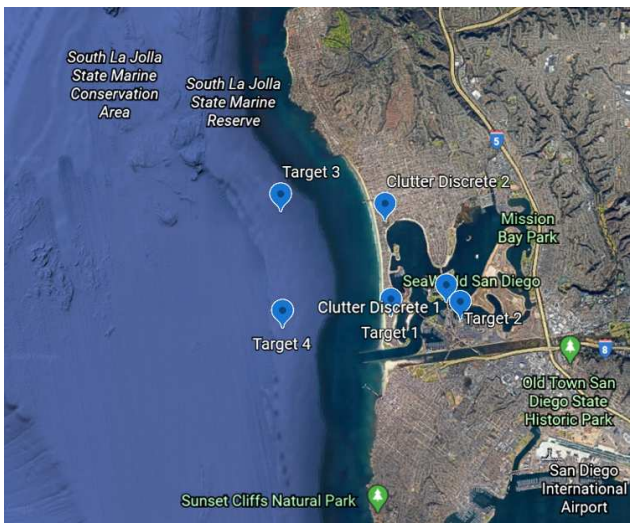}
    \caption{Scenario $1$ of the challenge dataset with $4$ targets and $2$ clutter discretes.}
    \label{scen1}
    \end{figure}

Along with the simulated data for this scenario, we have provided the true channel impulse responses for clutter and targets. This data spans $30$ CPIs, $32$ spatial channels, $64$ pulses, and $2334$ range bins. Basic beamforming and delay-Doppler processing of the data cube gives $30$ range-Doppler plots, one for each CPI. Along with the data, a reference video containing $30$ frames is also provided. Note that we used standard delay-Doppler processing and also assumed that a fixed waveform was transmitted. The goal is for the readers to test their own algorithms and optimally designed waveforms to improve target detection and estimation performance and obtain better results than the plots we demonstrate here with basic signal processing. The details of all the parameters chosen for this simulation are described in a user guide provided along with the challenge dataset.

\begin{figure}[htbp!]
    \centering
    \includegraphics[scale=0.75]{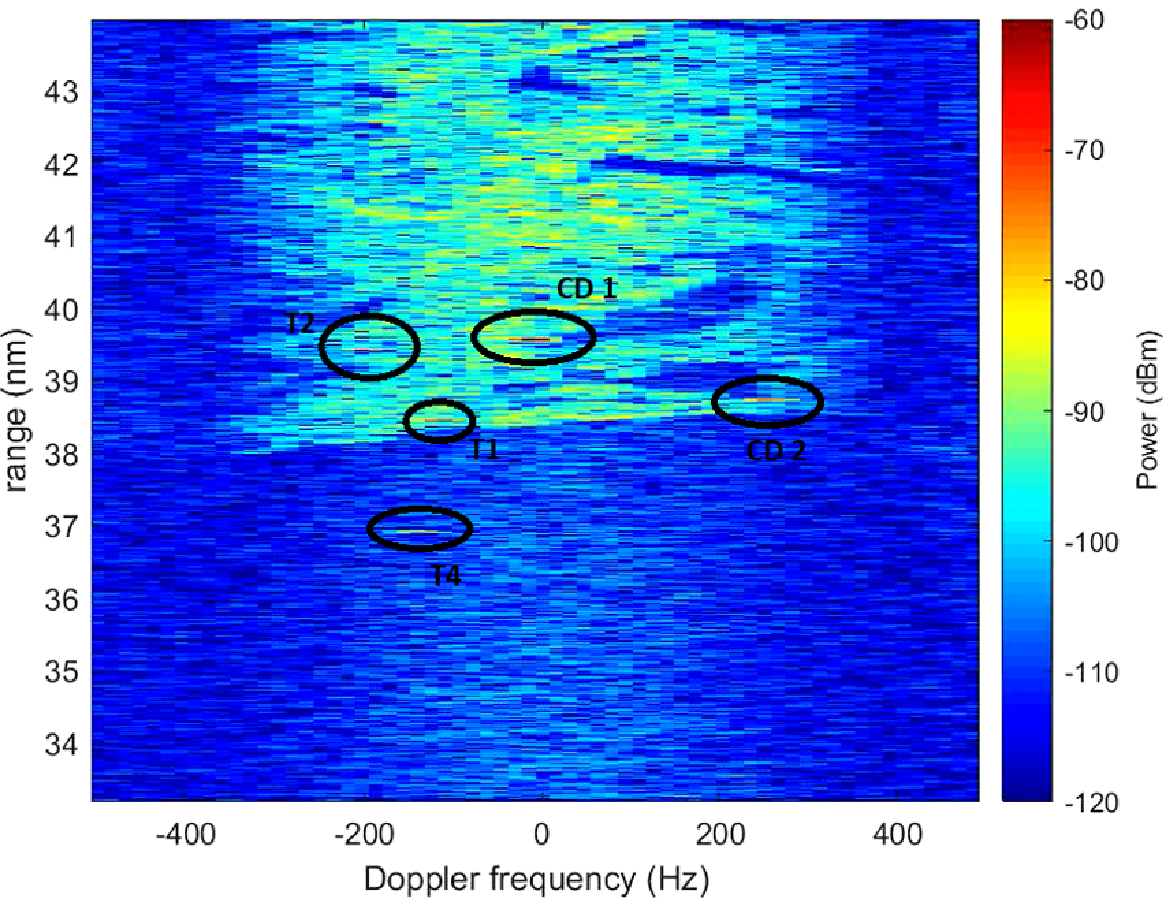}
    \caption{Range-Doppler plot from the $6$th CPI in scenario $1$.}
    \label{6}
    \end{figure}

\begin{figure}[htbp!]
    \centering
    \includegraphics[scale=0.75]{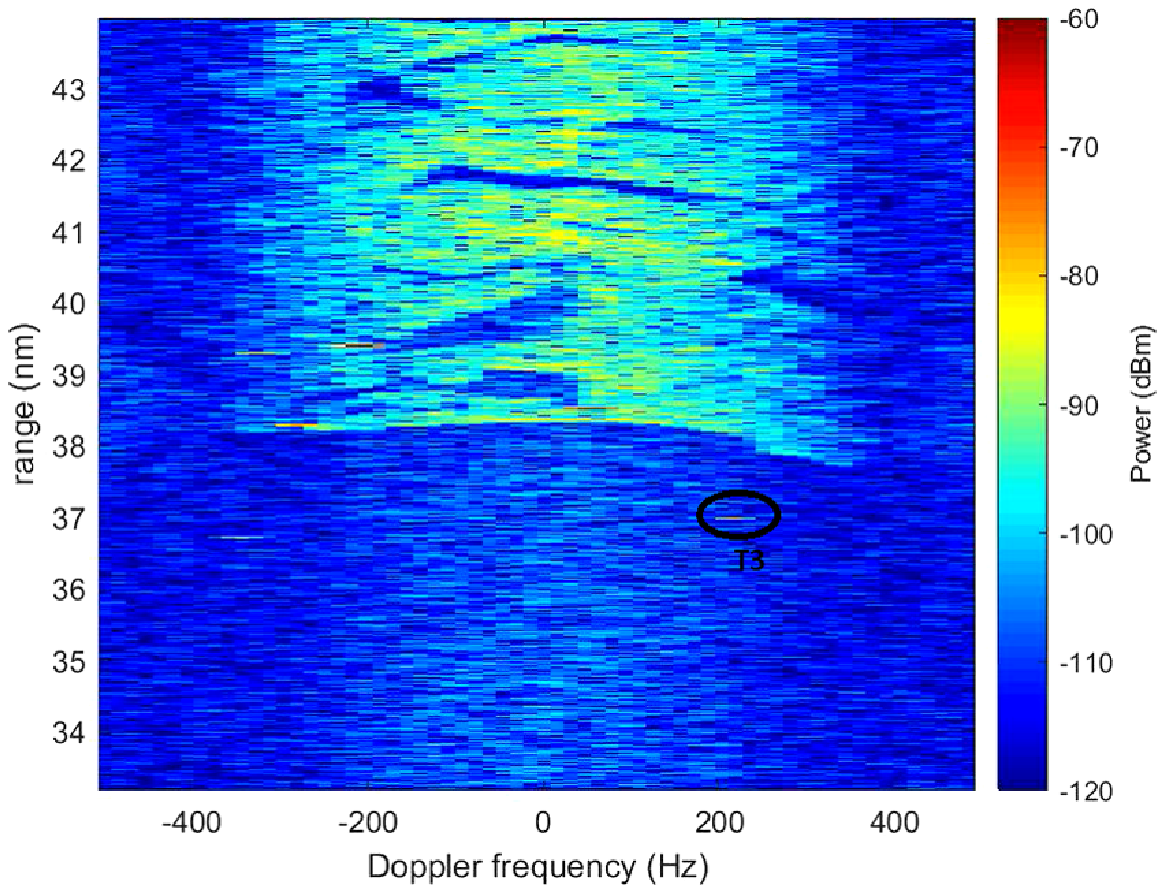}
    \caption{Range-Doppler plot from the $26$th CPI in scenario $1$.}
    \label{26}
    \end{figure}

For example, In the $6$th CPI, we obtain the plot in Fig. \ref{6}. Three targets and two clutter discretes can be identified from this plot. The other target which is much weaker and farther away from the other three targets is not visible in this plot. Also, from this plot, we can clearly observe littoral nature of the simulated environment as we can see regions of water within the ground clutter as water has weaker reflectivity compared to land. As the radar platform moves along and drags its beam along, we present the range-Doppler plot for the $26$th CPI in Fig. \ref{26}. Now, returns from the weaker target are also picked up by the radar. The other targets are still visible in the range-Doppler plot, but they have moved around from their positions in the previous plot (Fig. \ref{6}). Since the dataset also includes the true channel impulse responses corresponding to multiple pulses and CPIs, this dataset can be used to test fully adaptive radar optimal waveform design algorithms where waveforms are changed from pulse to pulse or from CPI to CPI. From the provided true channel impulse responses, data cubes can be generated for any transmit waveform using the model described in Section III.

\subsection{Scenario 2}
\begin{figure}[htbp!]
    \centering
    \includegraphics[scale=0.65]{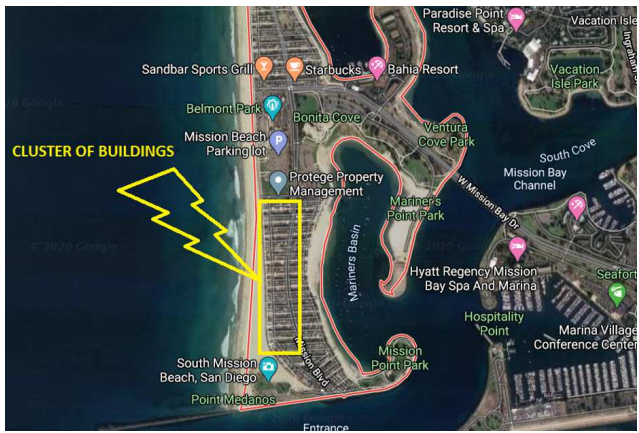}
    \caption{Cluster of buildings included as part of scenario $2$.}
    \label{scen2}
    \end{figure}

In scenario $1$, while we had ground clutter and couple of strong clutter discretes, they were spaced relatively far from the targets of interest. Hence, the detection and estimation of targets is not so difficult. Now, we move along to a more challenging dataset. In addition to the targets and strong clutter discretes present in scenario $1$, this scenario contains several ($150$) clutter discretes in the form of small buildings ($30\mathrm{m} \times 30\mathrm{m} \times 6\mathrm{m}$) arranged in a cluster very close to Target No. $1$. This makes this scenario more challenging than scenario $1$. The target locations and radar parameters remain the same as scenario $1$. $150$ buildings were added next to Target No. $1$ arranged in a $50\times 3$ grid. This is indeed a region populated by buildings as can be seen in satellite images (See Fig. \ref{scen2}). While all the buildings in this simulation were approximated to be of the same size, in more advanced datasets, each building can be modeled to be of the exact shape and size as in reality.

\begin{figure}[htbp!]
    \centering
    \includegraphics[scale=0.8]{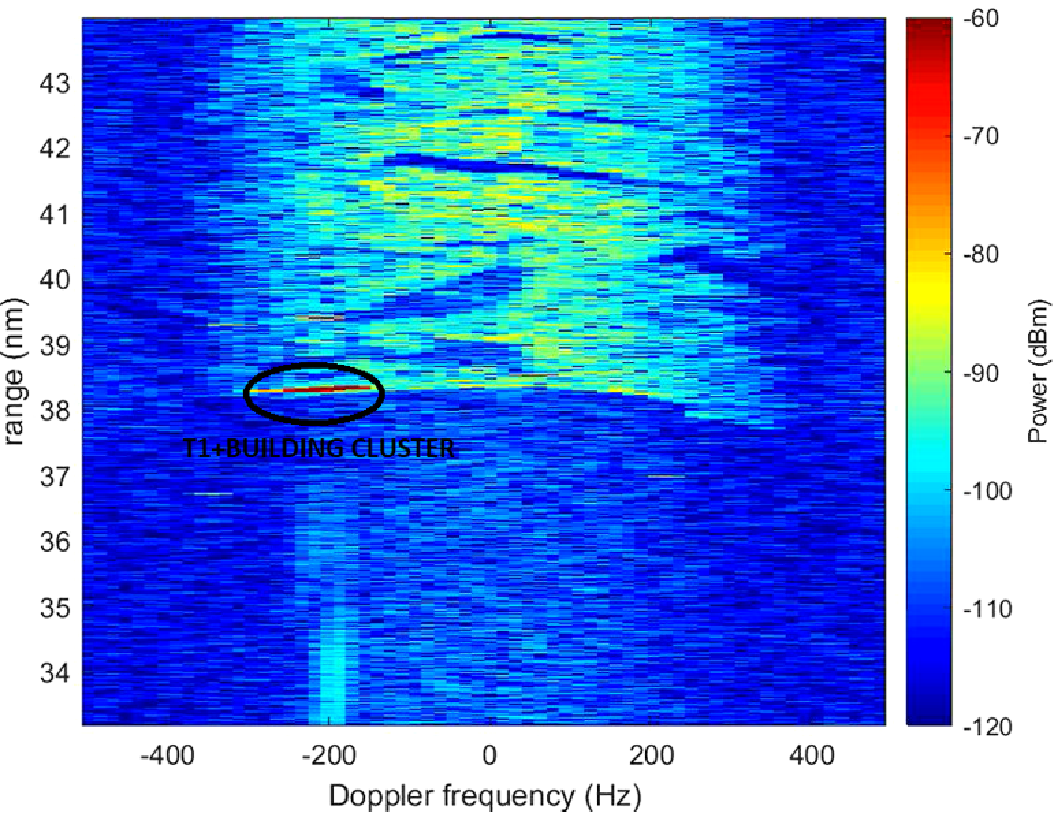}
    \caption{Range-Doppler plot from the $26$th CPI in scenario $2$.}
    \label{26-2}
    \end{figure}

In Fig. \ref{26-2}, we have the range-Doppler plot for the $26$th CPI in scenario $2$. We observe that the cluster of buildings is much stronger compared to Target $1$ and the target is not clearly distinguishable from the cluster of clutter discretes. Note that we have added this cluster of buildings all along the road on which this target was moving. This is a challenging scenario where CoFAR techniques for waveform design are needed to suppress the dominating clutter discretes. It is our hope that users of this dataset will devise advanced CoFAR techniques that can mitigate the effects of these clutter discretes using adaptive signal processing techniques and optimal waveform design.

\section{Concluding Remarks}
In this paper, we have reviewed advanced modeling and simulation techniques for modeling ground clutter using a stochastic transfer function approach. This approach is contrary to the traditional covariance based techniques and it lends itself well to accurately simulate various radar scenarios and applications as demonstrated in this paper. Lastly, we describe a new CoFAR challenge dataset that users can download, test, and benchmark state-of-the-art cognitive radar algorithms and techniques. It is our endeavour to generate and provide more advanced datasets involving realistic buildings and other cultural features.

\bibliographystyle{IEEEtran}
\bibliography{IEEEabrv,Sandeep_Ref}
\end{document}